%% file: main-rev1.tex
\documentclass[journal,twoside]{IEEEtran}
\IEEEoverridecommandlockouts
\usepackage{cite}
\usepackage{amsmath,amssymb,amsfonts,color}
\usepackage{algorithm,algpseudocode}
\usepackage{algorithmicx}
\usepackage{graphicx}
\usepackage{textcomp}
\usepackage{xcolor,soul}
\usepackage{tikz}
\usepackage{stfloats}
\usepackage{url}
\usepackage{array,ragged2e}
\usepackage[bookmarks=false]{hyperref}
\usepackage{trimclip} 

\ifCLASSOPTIONcompsoc
    \usepackage[caption=false, font=normalsize, labelfont=sf, textfont=sf, subrefformat=parens, labelformat=parens]{subfig}
\else
\usepackage[caption=false, font=footnotesize, subrefformat=parens, labelformat=parens]{subfig}
\fi

\def\BibTeX{{\rm B\kern-.05em{\sc i\kern-.025em b}\kern-.08em
    T\kern-.1667em\lower.7ex\hbox{E}\kern-.125emX}}

\newcommand{\wt}{\widetilde}
\newcommand{\wh}{\widehat}
\newcommand{\ds}{\displaystyle}

\usepackage[nodisplayskipstretch]{setspace}
\algnewcommand{\IfThenElse}[3]{
  \State \algorithmicif\ #1\ \algorithmicthen\ #2\ \algorithmicelse\ #3}

\algnewcommand{\IfThen}[2]{
  \State \algorithmicif\ #1\ \algorithmicthen\ #2}  

\input{my-macro.tex}

\begin{document}
	
\bstctlcite{IEEE_nodash:BSTcontrol}
\bstctlcite{IEEEexample:BSTcontrol}

\title{Approaching Massive MIMO Performance with Reconfigurable Intelligent Surfaces: \\ We Do Not Need Many Antennas}
\author{Giovanni Interdonato, \IEEEmembership{Member, IEEE,} Francesca Di Murro,  Carmen D'Andrea, \IEEEmembership{Member, IEEE,} Giovanni Di Gennaro, and Stefano Buzzi, \IEEEmembership{Senior Member, IEEE}%
\thanks{This paper was supported by the European Union under the Italian National Recovery and Resilience Plan (NRRP) of NextGenerationEU, partnership on “Telecommunications of the Future” (PE00000001 - program “RESTART”, Structural Projects 6GWINET and SRE, cascade call SPARKS).}
\thanks{A preliminary and simplified version of this article has been published in the proceedings of the 2021 ITG Workshop on Smart Antennas (WSA)~\cite{RISaidedWSA}.}
\thanks{The authors are with the Consorzio Nazionale Interuniversitario per le Telecomunicazioni (CNIT), 43124, Parma, Italy. G. Interdonato, C. D'Andrea and S. Buzzi are also with the Department of Electrical and Information Engineering (DIEI) of the University of Cassino and Southern Lazio, 03043 Cassino, Italy. 
G. Di Gennaro is also with Dipartimento di Ingegneria, Università degli Studi della Campania ``Luigi Vanvitelli'', Aversa (CE), 81031, Italy.
S. Buzzi is also affiliated with the Department of Electronics, Information and Bioengineering, Politecnico di Milano, 20133 Milano, Italy.}\vspace{-5mm}}

\maketitle

\begin{abstract}
This paper considers an antenna structure where a (non-large) array of radiating elements is placed at short distance in front of a reconfigurable intelligent surface (RIS), herein nicknamed reconfigurable intelligent base station (RIBS).
We firstly derive a closed-form expression {for the channel between the array of radiating elements and the RIS that captures the near-field effects,} and give some considerations on the channel hardening and favorable propagation in this scenario.
Focusing on both active and passive RIS, we describe channel estimation and downlink signal processing techniques suitable for the RIBS structure. Additionally, we formulate and solve an optimization problem aimed at maximizing the fairness among the users with respect to the downlink power coefficients and RIS configuration both in the cases of active and passive RIBS. 
Numerical results show that the proposed structure is effective and capable of outperforming conventional non-RIS aided MIMO systems, especially in the case of active RIBS. The proposed antenna structure is thus shown to be able to approach massive MIMO performance levels in a cost-effective way with reduced hardware resources.
\end{abstract}

\section{Introduction}

The increasing demand for ubiquitous connectivity along with a high \textit{quality of service} is driving a rapid technological (r)evolution of the mobile wireless systems.  
Massive multiple-input multiple-output (mMIMO) is undoubtedly a key physical layer technology for the current fifth generation (5G) of mobile wireless networks~\cite{redbook}. By leveraging the joint coherent transmission/reception from a large number of active antenna elements, and a fully digital baseband processing at the base station (BS), mMIMO enables an aggressive spatial multiplexing of a large number of user equipments (UEs) in the same time-frequency resource. This leads to unprecedented levels of coverage, spectral, and energy efficiency. Moreover, the use of large number of antennas triggers, in most of the propagation environments, pleasant phenomena deriving from the \textit{law of the large numbers} which simplify signal processing and resource allocation, thus reducing the hardware complexity and the circuit power consumption. Nevertheless, uncontrollably increasing the number of active antenna elements to improve the data rates is an expensive and not energy-efficient solution as the total energy consumption scales linearly with the number of radio frequency (RF) chains while the data rates grows logarithmically utmost~\cite{Ng2012}. 
Reconfigurable intelligent surfaces (RISs) constitute an emerging affordable solution to aid other technologies in implementing energy-efficient communications systems, and better coping with harsh propagation environments~\cite{Basar2019,Huang2019,WuQ2019}. An RIS is a meta-surface consisting of low-cost, typically passive, tiny reflecting elements that can be properly configured on real-time to either focus the energy towards areas where coverage or additional capacity is needed, or to null the interfering energy in specific spatial points. Basically, an RIS has the ability to shape the propagation channel and create a smart radio environment~\cite{DiRenzo2020} so as to improve the overall performance of the communication system. Technically, each RIS element introduces a configurable phase-shift to an impinging EM wave, so that the resulting reflected beam is steered towards the desired direction. Beamforming is therefore carried out by the environment in addition to the active radiative system where the impinging wave is generated. {Initially}, available RIS technology enabled the tuning of the phase-shift only of the reflected impinging waves, {and this kind of technology has been investigated in the literature to improve the performance of single-user and multiuser MIMO systems by finding the best phase configuration for the passive elements of the RIS\cite{Yan_JSAC2020,Ma_COMML2020,Liu_TVT2021}. Subsequently}, active RISs  have been introduced, with the capability of controlling both the amplitude and the phase of the reflected waves\footnote{Notice that the amplification is realized through  simple reflection-type amplifiers,  and no RF chains are needed (see \cite{ZhangZ2023} for further details).}~\cite{Long2021,ZhangZ2023}. 
In this paper we investigate a system where a nonconventional BS which consists of a planar antenna array illuminating a built-in RIS serves multiple UEs in a cellular setup. Such a structure is herein coined as \textit{reconfigurable intelligent base station} (RIBS). The goal of the paper is to show indeed that thanks to the use of the RIBS the multiplexing capabilities and performance of conventional MIMO systems can be approached with a much smaller number of active antennas and RF chains.

The architecture considered in this paper was firstly designed in the conference paper~\cite{RISaidedWSA} where we considered uniform linear array (ULA) BS and passive RIS and showed that the performance of mMIMO can be approached with a reduced number of active antennas. A similar architecture is also considered in~\cite{Mishra2023} where the use of a beyond-diagonal RIS is investigated.
Active RIS, capable of introducing a tunable amplification factor to reflected waves, are introduced in~\cite{Long2021,ZhangZ2023}, where it is discussed how the capacity gains of a passive RIS compared to mMIMO system become negligible in scenarios where the direct paths are heavily obstructed. A fair comparison between active and passive RIS with the same overall power budget is considered in reference~\cite{ZhiK2022}. The authors derive the optimal power splitting between the BS’s transmit signal power and active RIS’s output signal power and shows that the active RIS would be superior when the power budget is significant and the number of RIS elements is not very large. Active RISs are currently considered in many applications in the literature, some examples are in references \cite{LvW2023,Niu2023,ZhangJ2023}.
A novel stacked intelligent metasurface (SIM) architecture is proposed in~\cite{an2023stacked}  consisting of a multilayer metasurface structure  deployed at the BS to facilitate transmit beamforming and eliminating the need for conventional digital beamforming and high-resolution digital-to-analog converters at the BS.
Most of existing works focused on the joint optimization of the BS precoding vectors and the RIS phase-shifts, and propose effective channel estimation techniques to properly support this joint beamforming~\cite{Huang2019,WuQ2019,DiRenzo2020,WuQ2020}.
Reference~\cite{ZhiK2021} investigates the ergodic uplink (UL) rate of an RIS-aided system with direct links, assuming maximum-ratio combining (MRC) and zero-forcing (ZF) detector, respectively, with a low-overhead statistical channel state information (CSI)-based design for the RIS passive beamforming. 
Along this line,~\cite{Demir2021} proposes three architectures in which either a short-term RIS reconfiguration based on instantaneous CSI or a long-term RIS reconfiguration based on statistical CSI is considered.
In~\cite{Chen2021} a method to estimate the path gain and the angles of arrival is implemented by using limited RIS RF chains and pilots, assuming that some RIS elements are active. 
The work~\cite{Abrardo2021} presents a channel estimation scheme that capitalizes on the statistical characterization of the locations of the UEs for relaxing the need of frequently re-configuring the RIS phase-shifts. 
The problem of channel estimation in RIS-aided systems is also tackled in references~\cite{Chen2021,Araujo2021,Buzzi_TCCN_2021}.

Wireless communication systems have almost exclusively operated in the far field of antennas and antenna arrays, which is conventionally characterized by having propagation distances beyond the Fraunhofer distance. With the advent of extremely large aperture arrays and holographic mMIMO, the receiver will typically be in the \textit{near field} of the transmitter~\cite{Bjornson2020NearField}. Near-field communications have attracted substantial attention, especially in RIS-aided scenarios where the transmitter or receiver, or even both, operate in the near-field region of the RIS. Hence, the far-field assumptions are no longer applicable for the channel modeling and performance optimization. 
In \cite{TiwariCaire22} and \cite{TiwariCaire23} the authors focus on the properties of the near-field over-the-air propagation matrix between a small active multi-antenna feeder and a large RIS, both configured as standard linear arrays and placed in the near field of each other. In \cite{Cheng2024} an achievable rate optimization of a RIS-aided near-field wideband system is proposed, and in~\cite{Dovelos2021} the authors analytically determine the RIS response in the Fresnel zone by leveraging electromagnetic theory. Lastly,~\cite{Cui2022} studies the resonance to near-field channel estimation for large arrays.
\vspace{-0.3cm}
\subsection{{Motivation} and Contribution} \label{Motivation_and_Contribution_Subsection}
We consider an antenna structure where a (non-large) array of radiating elements is placed at short distance in front of an RIS, here called RIBS. 
{The motivation in considering this kind of architecture to emulate 
a conventional mMIMO system is given by the technical features that undoubtedly make the practical implementation of
such systems a challenging task. In particular, realizing transceivers with many active antenna elements, and thus RF chains, being able to simultaneously serve multiple users and performing digital beamforming, gives rise to a multitude of practical difficulties. For instance, the high fabrication cost, increased power consumption, constrained physical size and shape, and deployment limitations. As already discussed previously, metasurfaces offers a valid alternative to the conventional mMIMO system by providing additional degrees of freedom to improve the communications performance by exploiting less RF chains. The RIBS architecture considered in this paper at a glance can seem similar to dynamic metasurface antennas (DMAs)-aided transmitter \cite{DMA_WC2021,Yoo_DMA_TCOM2019,Shlezinger_DMA_TCOM2019}. DMAs consist of a multitude of reconfigurable
metamaterial radiating elements placed on a waveguide through which
the signals to be transmitted are transferred. The transceiver digital processor is connected to the waveguide
through dedicated input and output ports. The RIBS architecture discussed in this paper exploits a wireless channel between the BS and the RIS, thus the communication channel takes into account the different propagation scenario compared to DMAs-aided transmitter. Additionally, the RIBS architecture can be used to enhance the performance by properly integrating RISs on already existing BSs. The goal of investigating RIBS is to achieve mMIMO-like multiplexing gain and performance with lower hardware complexity.}

With reference to a single-cell system, we firstly derive a closed-form expression for the near field between the array of radiating elements and the RIS, then develop the signal model for both the cases of active and passive RIS, and provide theoretical considerations on the achievement of favorable propagation and channel hardening for the proposed RIBS structure. 
Then, we propose a joint optimization of RIS phase-shifts and downlink (DL) transmit powers, for both active and passive RIBS,  aimed at maximizing the fairness among the UEs. To address the non-convexity of such a joint optimization problem, we resort to alternate optimization programming. 
We perform extensive numerical simulations aimed at gaining insight into the performance of the proposed antenna architecture, also in comparison with conventional co-located mMIMO with active antennas and fully digital beamforming.
This paper demonstrates that RIBS can replicate conventional mMIMO performance with significantly less hardware complexity.

\subsection{{Paper Organization}}
The paper is organized as follows. Next section contains the description of the considered RIBS architecture, the channel model, and the channel estimation procedure. Section \ref{sec:performance-formulation} includes the discussion on the DL data transmission techniques, considerations on the performance measure {and the optimization problem formulation}. Section \ref{sec:optimization} is devoted to system optimization; in particular, this section addresses configuration of the active and passive RIBS and power allocation, including the power split between the active antenna array and the RIS (for the case of active RIBS). Section \ref{sec:num_results} discusses the numerical results, while findings and conclusions are summarized in Section \ref{sec:conclusions}.

\subsection{Notation}\label{subsec:notation}
We use non-bold letters for scalars, $a$ and $A$, lowercase boldface letters, $\mathbf{a}$, for vectors and uppercase lowercase letters, $\mathbf{A}$, for matrices. The transpose, the inverse and the conjugate transpose of a matrix $\mathbf{A}$ are denoted by $\mathbf{A}\trans$, $\mathbf{A}^{-1}$ and $\mathbf{A}\herm$, respectively. The trace of the matrix $\mathbf{A}$ is denoted as tr$\left(\mathbf{A}\right)$. {The operator $\diag(\cdot)$ acts differently according to its argument: $\bA\!=\!\diag(\ba)$ yields a diagonal matrix with the elements of the vector $\ba$ on the diagonal. Conversely, $\ba \!=\!\diag(\bA)$ yields a vector given by the diagonal elements of $\bA$.} The $N$-dimensional identity matrix is denoted as $\mathbf{I}_N$, the $(N\!\times\!M)$-dimensional matrix with all zero entries is denoted as $\mathbf{0}_{N\! \times\! M}$ and $\mathbf{1}_{N \!\times\! M} $ denotes a $(N \times M)$-dimensional matrix with unit entries. 
The statistical expectation and variance operators are denoted as $\EX{\cdot}$ and $\text{var}\{\cdot\}$ respectively; $\mathcal{CN}\left(\mu,\sigma^2\right)$ denotes a complex circularly symmetric Gaussian random variable with mean $\mu$ and variance $\sigma^2$.
\begin{figure*}
    \centering
    \hspace{-5mm}\subfloat[An illustrative example of the RIBS.\label{subfig:RIBS}]{%
       \resizebox{.68\columnwidth}{!}{\input{fig/RIBS}}}
    \hfill
  \subfloat[Back view of the RIBS.\label{subfig:back_view}]{%
        \resizebox{.72\columnwidth}{!}{\input{fig/back_view}}}
    \hfill
  \subfloat[Side view of the RIBS.\label{subfig:side_view}]{%
        \resizebox{.58\columnwidth}{!}{\input{fig/side_view}}}
    \hfill
  \caption{\protect\subref{subfig:RIBS}~A UPA with $\Na$ active antennas is mounted at close distance from an RIS with $\Nr$ reflective elements. The relative positioning of the active array with respect to the RIS is such that no UE is obstructed. \protect\subref{subfig:back_view}~The BS antennas have size $\Dah$ along the $x$-axis and $\Dav$ along the $y$-axis, and are spaced by $\dah$ and $\dav$ in the $x$- and $y$-axis, respectively. The RIS elements have size $\Drh$ along the $x$-axis and $\Drv$ along the $y$-axis, and are spaced by $\drh$ and $\drv$ in the $x$- and $y$-axis, respectively. \protect\subref{subfig:side_view}~View of the RIBS in the $yz$-plane, and characterization of the cascaded channel. $D$ is the distance between the RIS and the BS, while $\alpha$ denotes the downtilt angle of the BS with respect to the RIS. The heights of RIS, BS and UE are denoted by $h_{\RIS}$, $h_{\BS}$ and $h_{\UE}$, respectively.}
  \vspace{-3mm}            
\end{figure*}

\section{System model}
Consider a single-cell network, operating in \textit{time division duplexing} (TDD) and at the sub-6 GHz frequencies, where $K$ single-antenna UEs are served by a RIBS depicted in Fig.~\ref{subfig:RIBS}.
We denote by $\Na \!=\! \Nah \Nav$ the elements of the planar antenna array, hereafter simply referred to as BS, with $\Nah$ and $\Nav$ being the elements along the horizontal and vertical axis, respectively. This BS transmits a signal that has polarization in the $y$-direction and travels along the $z$-direction. 
Similarly, we let $\Nr \!=\! \Nrh \Nrv$, with $\Nr\!>\!\Na$, be the number of configurable reflective elements of the RIS. The BS antenna elements have size $\Dah$ along the $x$-axis and $\Dav$ along the $y$-axis, and are spaced by $\dah$ and $\dav$ in the $x$- and $y$-axis, respectively. The RIS elements have size $\Drh$ along the $x$-axis and $\Drv$ along the $y$-axis, and are spaced by $\drh$ and $\drv$ in the $x$- and $y$-axis, respectively. \Figref{subfig:back_view} shows the back view of the RIBS in the $xy$-plane by emphasizing its geometry.
The active planar array (BS) is placed at a distance $D$ from the RIS and with a tilt angle $\alpha$ with respect to the RIS plane. The heights of RIS, BS and UE are denoted by $h_{\RIS}$, $h_{\BS}$ and $h_{\UE}$, respectively. \Figref{subfig:side_view} illustrates the side view of the RIBS in the $yz$-plane.
Let $i_h$ and $i_v$ be the indices uniquely identifying a RIS element along the horizontal and vertical axis, respectively. Similarly, let us introduce the indices  $j_h$ and $j_v$ to identify an antenna element of the BS. Then, the coordinates of an arbitrary RIS element $(i_h, i_v)$, in our 3D reference system, are given by
\begin{align}
C_\RIS(i_h, i_v) \!&=\! \left[(1\!-\!\Nrh)\drh/2 \!+\!(i_h\!-\!1)\drh, \right. \nonumber \\
&\qquad \left.(1\!-\Nrv\!)\drv/2\!+\!(i_v\!-\!1) \drv\!+\!h_\RIS,~0\right].
\end{align}  
Similarly, the coordinates of an arbitrary antenna element $(j_h, j_v)$ of the BS are given by 
\begin{align}
C_\BS(j_h, j_v) \!&=\! \left[(1-\!\Nah) \dah/2 \!+\!(j_h\!-\!1) \drh, \right. \nonumber \\
&\qquad \left. \left((1\!-\!\Nav)\dav/2\!+\!(j_v\!-\!1)\dav\right)\cos\alpha\!+\!h_\BS, \right. \nonumber \\
&\qquad \left. D\!-\!\left((1\!-\!\Nav)\dav/2\!+\!(j_v\!-\!1) \dav\right)\sin\alpha\right].
\end{align}
The RIS may be either \textit{passive} or \textit{reflective active}.\footnote{Transmissive and hybrid active RISs~\cite{Mu2021,Liu2021,Zeng2021} are out of scope.}
In the former, each element of the RIS solely introduces a tunable phase offset on the impinging waves, and does not alter its amplitude; whereas, in the latter, we assume that each element of the RIS performs an amplify-and-reflect operation, as described in~\cite{ZhangZ2023}, so that the reflected wave is modified both in its amplitude and phase. Such a reflective active RIS is obtained by exclusively integrating an active reflection-type amplifier in each RIS element, and differs from an active RIS equipped with an RF chain per active element. The latter needs higher hardware complexity to support its baseband processing capability, and operates similarly to a decode-and-forward relay~\cite{DiRenzo_OJCOMS2020,Nguyen2021,He2021}. Conversely an active reflection-type amplifier can be implemented by using low-cost, low-power-consumption components such as cross-coupled structures of transistors~\cite{Bousquet2012}, multi-layer integrated circuits~\cite{Kishor2012}, and current-inverting converters~\cite{Loncar2109}.
Regardless of the RIS typology, the effect of the RIS is modeled as a diagonal $(\Nr\! \times\! \Nr)$-dimensional matrix, denoted by $\bP$. The diagonal entries of $\bP$ are constrained to have unit modulus in case of passive RIS, whereas they have a tunable magnitude in case of active RIS. The RIS is controlled by the BS which is able to adjust its configuration (i.e., fine-tuning $\bP$) whenever required. 

\subsection{Channel Model}
\label{subsec:channel-model}
The conventional block-fading channel model is considered, and let $\tc$ denote the channel coherence block length in samples. In TDD mode, each coherence block accommodates UL training, UL and DL data transmission, such that $\tc\!=\!\tp\!+\!\tu\!+\!\td$, where $\tp$, $\tu$ and $\td$ are the training duration, the UL and the DL data transmission duration, respectively. 

Let $\bh_k$ be the $\Nr$-dimensional vector characterizing the UL channel from an arbitrary UE $k$ to the RIS, and $\bH$ be the $(\Na\!\times\!\Nr)$-dimensional channel vector from the RIS to the BS. Then, the cascaded UL channel from UE $k$ to the RIBS is the $\Na$-dimensional vector 
\begin{equation}
	\bar{\bh}_k = \bH \bP \bh_k = \bH \bH_k \bp = \bG_k \bp  \, ,
	\label{eq:composite_signature}
\end{equation}
where $\bp \!=\! \diag(\bP)\!=\![p_1,\ldots,p_{\Nr}]\trans$, $\bH_k \!=\!\diag(\bh_k) \!\in\!\mC^{\Nr\times\Nr}$, and we have defined $\bG_k \!\triangleq\! \bH \bH_k \!\in\! \mC^{\Na \times \Nr}$.
As per the channel between the UEs and the RIS, we assume spatially correlated Rayleigh fading~\cite{Bjornson2021Rayleigh}, $\bh_k \!\sim\!\CG{\bzero_{\Nr}}{\bR_k}$, where $\bR_k\!\in\!\mC^{\Nr\times\Nr}$ is the channel covariance matrix of UE $k$ that captures the effects of pathloss, shadowing and spatial correlations. Moreover, we let $\beta_k$ be the large-scale fading coefficient (i.e., the average channel gain) of UE $k$ obtained as $\beta_k \!=\! \tr(\bR_k)/\Nr$. All the channel statistics are conventionally assumed to be known at the BS in each coherence interval.
As per the channel between the RIS and the BS, we consider the following assumptions:
\begin{itemize}
	\item[i)] The BS does not cause any blockage on the electromagnetic radiation reflected by the RIS. In practical scenarios, the BS can be placed laterally with respect to the RIS so as to simply avoid the blockage problem. 
	\item[ii)] The UEs to be served are placed at the backside of the BS, thus the direct links from the BS to the UEs are neglected as being considerably weaker than those of the cascaded links through the RIS.
	\item[iii)] Given the short distance between the RIS and the BS, the channel $\bH$ is reasonably modeled as a deterministic quantity, and it is known at the BS.
	\item[iv)] The mutual coupling between the antenna elements of the BS and between the reflective elements of the RIS are neglected. This assumption is usually met provided that the element spacing at both the BS and the RIS is not smaller than $\lambda/2$~\cite{qian2021mutual}, with $\lambda$ being the wavelength.  
\end{itemize}

\subsubsection{Far-field Zone}\label{subsec:far-field}
We assume that each element of the RIS is in the far-field zone of each element of the BS. However, the RIS as a whole is not required to be in the far-field zone of the whole BS.
{In this regard,  let $d_{m,n}$ be the distance between the centers of the $m$-th BS antenna and the $n$-th RIS elements.} Then, the far-field condition among an arbitrary pair of BS antenna element $m$ and RIS element $n$ is  given by 
$d_{m,n} \!>\! d_{{\rm F}} =(2/\lambda)\max\{ \Da^2, \Dr^2\}$, and $d_{m,n} \!\gg\! \max\{ \Da, \Dr, \lambda\}$ with $\Da\!=\!\max\{ \Dah,\Dav \}$ and $\Dr\!=\!\max\{ \Drh,\Drv \}$~\cite{stutzman2012antenna}. 
The value $d_{{\rm F}}$ is also known as the Fraunhofer distance and it constitutes the limit between the near- and far-field zones.
Conversely, the far-field condition among the RIS and the BS is given by $D \!>\! (2/\lambda)\max\{ (\Na\da)^2, (\Nr\dr)^2\}$ and $
D \!\gg\! \max\{ \Na\da, \Nr\dr, \lambda\}$.
Under this condition and according to the standard electromagnetic radiation theory, the $(m,n)$-th entry of $\bH$, which denotes the UL channel coefficient from the $n$-th element of the RIS to the $m$-th BS antenna, can be written as~\cite{Jamali_Tulino_OJCOMs_2021}:
\begin{align}
	[\bH]_{m,n} \!\!&= \!\frac{\sqrt{\rho \lambda^2 G_\tA G_\tR}}{4 \pi d_{m,n}} \euler^{-\imagj 2 \pi d_{m,n}/\lambda} \!\prod\limits_{\ell=1}^2 {f(\theta_{\ell,m,n},\phi_{m,n})}\, ,
\label{eq:H_ij_planar}
\end{align}
where $\rho\leq1$ is a real-valued coefficient modeling the RIS efficiency in reflecting the impinging waves, ${f(a,b) = \cos a \cos b}$, and
\begin{align}
\begin{cases}
G_\tA \!&\!=\! (4 \pi/\lambda^2)A_{\text{effA}} \, , \\
A_{\text{effA}} \!&\!=\! \Dah \Dav \, ,
\end{cases} \quad
\begin{cases}
G_\tR \!&\!=\! (4 \pi/\lambda^2)A_{\text{effR}} \, , \\
A_{\text{effR}}\!&\!=\!\Drh \Drv \, ,
\end{cases}
\end{align}
{where $G_\tA$, $G_\tR$, $A_{\text{effA}}$, $A_{\text{effR}}$ are the antenna gains and the effective apertures of the BS and the RIS, respectively.}
Moreover, letting $y^\tC_{m,n}$ and $z^\tC_{m}$ be the distance between the centers of BS antenna element $m$ and RIS element $n$ along the $y$- and $z$-axis, respectively, then $\theta_{1,m,n} \! =\! \alpha-\arctan\left({y^\tC_{m,n}}/{z^\tC_{m}}\right)$ and $\theta_{2,m,n}\!=\! {\pi}/{2}-\arctan\left({z^\tC_{m}}/{y^\tC_{m,n}}\right)$
are the azimuth angles, while the elevation angles are given by $\phi_{m,n} \!=\! \arctan({y^\tC_{m,n}}/{D})$.
%
\subsubsection{Radiative Near-field Zone} \label{subsec:near-field}
We now present a characterization of the channel between the RIS and the BS in the radiative near-field zone.
In the radiative near field, namely wherever $d_{\text{N}} \!<\! d_{m,n}\!<\! d_{\text{F}}$, with $d_{\text{N}}$ being the distance at which the \textit{reactive} near field ends, the free-space normalized channel gain between the $m$-th antenna element of the BS, with spatial coordinates $(x^\tA_{m,n}, y^\tA_{m,n}, z^\tA_{m,n})$, and the $n$-th element of the RIS, with spatial coordinates $(x^\tR_{m,n}, y^\tR_{m,n}, z^\tR_{m,n})$, is upper bounded, by using the Cauchy-Schwarz inequality~\cite[Appendix A]{Bjornson2020NearField}, as shown in~\eqref{eq:channel-gain:near-field} at the top of next page, %
where $1\!<\!n\!<\!\Nr$, $1\!<\!m\!<\!\Na$, the functions $F_1$, $F_2$, $F_3$ and $F_4$ are defined in~\eqrefs{eq:F1}{eq:F4}, and $r$ is the Euclidean norm of an arbitrary point with spatial coordinates $(x,y,z)$ located in the Fraunhofer radiation region of the source.
\begin{figure*}[!t]
\small
\begin{align}
\!\!\!\!\!\left|[\bH]_{m,n}\right|^2 & \!  \leq  \! \Bigg[\sum_{x\in \setX_{1,m,n}} \sum_{y\in \setY_{1,m,n}} \!\!F_{1}(x,y,y_{1,m,n},x_{1,m,n},z) \!+\!\!\!\sum_{x\in \setX_{1,m,n}} \sum_{y \in \setY_{2,m,n}}  \!\!F_2(x,y,x_{1,m,n},z) \!+\!\!\!\sum_{x\in \setX_{2,m,n}} \sum_{y \in \setY_{2,m,n}} \!\!F_4(x,y,z)\Bigg. 	\nonumber\\
&\qquad+\!\!\!\sum_{x\in \setX_{2,m,n}} \sum_{y \in \setY_{1,m,n}} \!\!F_3(x,y,y_{1,m,n},z) \!-\!\sum_{x\in \setX_{1,m,n}} \sum_{y \in \setY_{3,m,n}} \!\!F_1(x,y,y_{4,m,n},x_{1,m,n},z)\nonumber \\
&\qquad-\!\!\!\sum_{x\in \setX_{3,m,n}} \sum_{y\in \setY_{1,m,n}} \!\!F_1(x,y,y_{1,m,n},x_{4,m,n},z)\!+\!\sum_{x\in \setX_{3,m,n}} \sum_{y\in \setY_{3,m,n}} \!\!F_1(x,y,y_{4,m,n},x_{4,m,n},z)\nonumber \\[-0.5ex]
&\Bigg.\qquad-\!\!\!\sum_{x\in \setX_{2,m,n}} \sum_{y\in \setY_{3,m,n}} \!\!F_3(x,y,y_{4,m,n},z) \!-\!\sum_{x\in \setX_{3,m,n}} \sum_{y\in \setY_{2,m,n}} \!\!F_2(x,y,x_{4,m,n},z)\Bigg] \Delta_{m,n},
\label{eq:channel-gain:near-field}\\[.5ex]
F_{1}(x,y,\bar{y},\bar{x},z) &= -\frac{\bar{x}}{3z^{2}}\left[\frac{2xz^{2}(x^{2}+y\bar{y}+z^{2})}{(x^{2}+z^{2})r}+\left(\bar{y}\sqrt{-z^{2}}-z^{2}\right)\tanh^{-1} \left(\frac{x^{2}-y\sqrt{-z^{2}}+z^{2}}{xr}\right) +x\bar{x}\frac{x^{2}+y\bar{y}+z^{2}}{(x^{2}+z^{2})r} \right. \nonumber \\ 
&\qquad-\left.\left(\bar{y}\sqrt{-z^{2}}+z^{2}\right)\tanh^{-1}\left(\frac{x^{2}+y\sqrt{-z^{2}}+z^{2}}{xr}\right) \right] + \frac{1}{r(x^{2}\!+\!z^{2})} \nonumber\\
&\qquad\!\times\!\Bigg[2x^{4}\!+\!x^{2}(y^{2}\!+\!y\bar{y}\!+\!3z^{2})\!-\!\bar{y}(x^{2}\!+\!z^{2})^{\frac{3}{2}}\left(\frac{r^{2}}{x^{2}\!+\!z^{2}}\right)^{\frac{1}{2}} \log\left(\sqrt{\frac{y^{2}}{(x^2\!+\!z^2)}}\!+\!\frac{y}{\sqrt{x^{2}\!+\!z^{2}}}\right) \!+\! z^{2}(y^{2}\!+\!z^{2})\Bigg] \, , \label{eq:F1}\\
F_{2}(x,y,\bar{x},z) &=  -\frac{2x^{3}\bar{x}}{3} \Delta^\ty\left[\arctan\left(\frac{z\sqrt{y^{2}+z^{2}+x^{2}}}{x|y|}\right)\frac{1}{zx^{3}} +\frac{|y|}{(x^{2}+z^{2})rx^{2}}\right]-\frac{x\bar{x}y \Delta^\ty}{(x^{2}+z^{2})r}  \nonumber \\
 &\qquad- \frac{(x^{2}+z^{2})^{\frac{3}{2}}\sqrt{\dfrac{r^2}{z^{2}+x^{2}}}\sinh^{-1}\left(\dfrac{y}{x^{2}+z^{2}}-x^{2}y\right)}{(x^{2}+z^{2})3r} \Delta^\ty \, ,\\
F_{3}(x,y,\bar{y},z) &= \Bigg[ \frac{2xz^{2}(x^{2}+y\bar{y}+z^{2})}{3z^{2}(x^{2}+z^{2})r} + \left(\bar{y}\sqrt{-z^{2}}-z^{2}\right)x\tanh^{-1} \left(\frac{x^{2}-y\sqrt{-z^{2}}+z^{2}}{xr}\right)\nonumber \\ 	
&\qquad\Bigg. -\left(\bar{y}\sqrt{-z^{2}}+z^{2}\right)x\tanh^{-1} \left(\frac{x^{2}+y\sqrt{-z^{2}}+z^{2}}{xr}\right)\Bigg]\Delta^\tx \, ,\\
F_{4}(x,y,z) &= \left[ \frac{2x^{3}}{3}\arctan\left(\frac{z\sqrt{y^{2}+z^{2}+x^{2}}}{x|y|}\right)\frac{1}{zx^{3}} +\frac{x|y|}{(x^{2}+z^{2})rx^{2}}\right]\Delta^\ty\Delta^\tx \, .
\label{eq:F4}
\end{align}
\hrulefill
\vspace*{-3mm}
\end{figure*}
In addition, $\Delta_{m,n}=\Drh \, \Drv \, \Delta^\tx_{m,n}\, \Delta^\ty_{m,n}$, where $\Delta^\tx_{m,n} \!=\! A^\tx_{m,n}\!-\!B^\tx_{m,n}$ and $\Delta^\ty_{m,n} \!=\! C^\ty_{m,n}\!-\!B^\ty_{m,n}$.
We also define $\setX_{1,m,n}\!=\![x_{2,m,n},-x_{1,m,n}]\, ,\quad 
 \setY_{1,m,n}\!=\![y_{2,m,n},-y_{1,m,n}]\, ,
\setX_{2,m,n}\!=\![x_{3,m,n},-x_{2,m,n}]\, ,\quad 
\setY_{2,m,n}\!=\![y_{3,m,n},-y_{2,m,n}]\, ,
\setX_{3,m,n}\!=\![x_{4,m,n},-x_{3,m,n}]\, ,\quad 
\setY_{3,m,n}\!=\![y_{4,m,n},-y_{3,m,n}]\, ,$ and, 
%
\begin{align*}
	x_{1,m,n} &\!=\! x^\tR_{m,n}\!-\!\frac{\Drh}{2}\!-\!A^\tx_{m,n}\, ,\quad
	y_{1,m,n} \!=\! y^\tR_{m,n}\!-\!\frac{\Drv}{2}\!-\!C^\ty_{m,n}\, , \\
	x_{2,m,n} &\!=\! x^\tR_{m,n}\!-\!\frac{\Drh}{2}\!-\!B^\tx_{m,n}\, ,\quad
	y_{2,m,n} \!=\! y^\tR_{m,n}\!-\!\frac{\Drv}{2}\!-\!B^\ty_{m,n}\, , \\
	x_{3,m,n} &\!=\! x^\tR_{m,n}\!+\!\frac{\Drh}{2}\!-\!A^\tx_{m,n}\, ,\quad 
	y_{3,m,n} \!=\! y^\tR_{m,n}\!+\!\frac{\Drv}{2}\!-\!C^\ty_{m,n}\, , \\
	x_{4,m,n} &\!=\! x^\tR_{m,n}\!+\!\frac{\Drh}{2}\!-\!B^\tx_{m,n}\, ,\quad 
	y_{4,m,n} \!=\! y^\tR_{m,n}\!+\!\frac{\Drv}{2}\!-\!B^\ty_{m,n}\, , 
\end{align*}
{The quantities $A^\tx_{m,n}$, $B^\tx_{m,n}$ and $C^\ty_{m,n}$, $B^\ty_{m,n}$ are the integration extremes related to the position of the BS tilted with respect to the RIS as shown in the Appendix ~\ref{subsec:appendix-nearfield}. Specifically, $A^\tx_{m,n}$, $B^\tx_{m,n}$ being the upper and lower extreme, respectively, of the $m$-th BS antenna in the {$xz$}-plane, and $C^\ty_{m,n}$, $B^\ty_{m,n}$ being the upper and lower extreme, respectively, of the $m$-th BS antenna in the {$yz$}-plane, with respect to the reference point centered at the $n$-th RIS element. These positions have been calculated starting from geometric considerations that mainly include notions of trigonometry. We report their expressions in the Appendix ~\ref{subsec:appendix-nearfield}.}


%
{The right-hand side in~\eqref{eq:channel-gain:near-field} is obtained by computing a 4-D integral which, through mathematical manipulations, is transformed into the sum of nine double integrals, reported in Appendix~\ref{subsec:appendix-nearfield}, each of which can be calculated analytically. For the sake of brevity, we here detail the calculation of the first of these nine double integrals by omitting the indices $m$ and $n$ identifying the BS antenna and the RIS element, respectively. The double integral to be computed is
\begin{equation}
\label{eq:IF1}
\!\!\!\!I_1 \!=\!\! \int_{x_1}^{x_2}\!\!\int_{y_1}^{y_2}\!\!\!(x-x_1) (y-y_1) \frac{y^2+z^2}{(x^2+y^2+z^2)^ \frac{5}{2}}dx\,dy\, .
\end{equation}%
By solving \eqref{eq:IF1} first with respect to the variable $x$, we obtain: 
\begin{equation}
\label{eq:IF1_2}
\!\!\!\!I_1 \!=\!\! \sum_{x\in \setX_1}\!\!\int_{y_1}^{y_2}\!\!\!\!(y-y_1) \!\!\left[\frac{-2x^3x_1\!-\!3xx_1(y^2+z^2)\!-\!(y^2\!+\!z^2)}{3(y^2\!+\!z^2)(x^2\!+\!y^2\!+\!z^2)^\frac{3}{2}}\right]\! dy\, ,
\end{equation}
then, by solving~\eqref{eq:IF1_2} with respect to the variable $y$, we obtain 
\begin{equation}
\label{eq:IF1_3}
\!\!\!I_1 \!=\!\! \sum_{x\in \setX_1}\sum_{y\in \setY_1} F_1(x,y,x_1,y_1,z) \, ,
\end{equation}%
where $F_1(x,y,x_1,y_1,z)$  is defined in \eqref{eq:F1}.}
The $(m,n)$-th entry of the RIS-to-BS $(\Na \!\times\! \Nr)$-dimensional matrix $\bH$ is 
\begin{equation}
[\bH]_{m,n} = |[\bH]_{m,n}| \exp(-\imagj 2\pi d_{m,n}/\lambda)\, .
\label{eq:Hnear}
\end{equation}
\begin{IEEEproof}
See Appendix~\ref{subsec:appendix-nearfield}.
\end{IEEEproof}

\subsection{Favorable Propagation and Channel Hardening}
Before proceeding to the system design and analysis, some considerations are needed about the channel hardening and favorable propagation. 
First of all, we note that, given the UE $k$ composite signature in \eqref{eq:composite_signature}, the case of a traditional MIMO array with $\Nr \!=\! \Na$ active antennas can be obtained by letting $\bH\!=\!\bP\!=\!\bI_{\Na}$. For such a system, as it is well known, under the assumption of uncorrelated fast fading realizations across antennas, both \textit{favorable propagation} and \textit{channel hardening} phenomena are observed increasing values of $\Na$. Favorable propagation refers to the fact that the inner product between two different channel signatures converges to zero almost surely as the number of antennas diverges, while channel hardening refers to the fact that the squared magnitudes of the composite signatures converge to a deterministic value (that is the corresponding large scale fading coefficient multiplied by the number of antennas) as the antenna array size grows large. For finite values of the number of antennas, the favorable propagation condition can be checked by verifying that 
\begin{equation}
f_{k,j} \triangleq \text{var}\left\{\frac{\bar{\bh}_k\herm \bar{\bh}_j}{
\ds \sqrt{\EX{\|\bar{\bh}_k\|^2 }\EX{\|\bar{\bh}_j\|^2 }}
} \right\} \rightarrow 0 
 \; ,
\label{eq:favourable}
\end{equation}
for all $k \neq j$, when the number of antennas grow large, 
while the channel hardening condition is checked by verifying that
\begin{equation}
f_{k,k}\triangleq \text{var}\left\{\|\bar{\bh}_k\|^2/
	\ds \EX{\|\bar{\bh}_k\|^2 } \right\} \rightarrow 0 \; , 
\label{eq:hardening}
\end{equation}
for all $k$, again in the large array limiting regime. Specifically, it is easy to show that for a conventional mMIMO system (i.e., with no RIS) with $\Na$ antennas, both \eqref{eq:favourable} and \eqref{eq:hardening} are proportional to $1/\Na$, and vanish as $\Na \rightarrow +\infty$. As for the proposed RIBS architecture, letting $\wt{\bR}_{k,\bP}= \bH \bP \bR_k \bP\herm \bH\herm \!\in\!\mC^{\Na\!\times\!\Na}$, the term $f_{k,j}$ is given in closed form as
\begin{equation}
f_{k,j} \!=\! \frac{\tr\bigl( \wt{\bR}_{k,\bP} \wt{\bR}_{j,\bP}\bigr)}{
\tr\bigl( \wt{\bR}_{k,\bP}\bigr) \tr\bigl( \wt{\bR}_{j,\bP}\bigr)}  \!=\! \frac{\sum_{\ell=1}^{\Na}\wt{\lambda}_\ell^{(k,j)}}{\sum_{\ell=1}^{\Na}\lambda_\ell^{(k)}\sum_{\ell=1}^{\Na}\lambda_\ell^{(j)}}\,, 
\label{eq:fkj}
\end{equation}
where $\wt{\lambda}_1^{(k,j)}, \ldots, \wt{\lambda}_{\Na}^{(k,j)}$, $\lambda_1^{(k)}, \ldots, {\lambda}_{\Na}^{(k)}$ and $\lambda_1^{(j)}, \ldots, {\lambda}_{\Na}^{(j)}$ are the non-zero eigenvalues of the matrices $\wt{\bR}_{k,\bP} \wt{\bR}_{j,\bP}$, $\wt{\bR}_{k,\bP}$ and $\wt{\bR}_{j,\bP}$, respectively. 
\begin{lemma}\label{lemma:Fp-Ch}
As for the RIBS setup, $f_{k,k}$ and $f_{k,j}$ defined in \eqref{eq:favourable} and \eqref{eq:hardening}, respectively, scale with $1/\Na$.
\end{lemma}%
\begin{IEEEproof}
    The proof readily follows by finding {an upper bound of~\eqref{eq:fkj} and by using the maximum eigenvalue of matrix $\wt{\bR}_{k,\bP} \wt{\bR}_{j,\bP}$, $\wt{\lambda}_{{\rm max}}^{(k,j)}$, and the minimum eigenvalues of the matrices $\wt{\bR}_{k,\bP}$ and $\wt{\bR}_{j,\bP}$, $\lambda_{\rm min}^{(k)}$ and $\lambda_{\rm min}^{(j)}$, respectively, 
    \begin{equation}
    f_{k,j}\leq \frac{1}{\Na}\frac{\wt{\lambda}_{{\rm max}}^{(k,j)}}{\lambda_{\rm min}^{(k)}\lambda_{\rm min}^{(j)}} \propto \frac{1}{\Na}.
    \label{eq:fkj2_LB}
    \end{equation}}
    The proof concerning $f_{k,k}$ follows the same considerations.
\end{IEEEproof}

Lemma~\ref{lemma:Fp-Ch} reveals a somewhat disappointing result: the favorable propagation and channel hardening are not affected by the size of the RIS, but rather depend on the number of BS antennas. It could thus appear that there is no advantage in considering the proposed RIBS in comparison to a co-located mMIMO BS array with $\Na$ antennas. A moment thought clarifies however that the Lemma just suggests that the statistical parameters ruling channel hardening and favorable propagation do not benefit from the use of an RIS. The additional degrees of freedom provided by the tunable RIS elements can be however exploited to improve the performance.

\subsection{UL Training}
\label{subsec:uplink-training}
The UL training phase consists of $\Nr\!+\!1$ different pilot transmission intervals, each spanning $\tp \!=\! K$ channel uses. In each interval, UE $k$ transmits a known pilot sequence, denoted by $\bvphi_k \!\in\! \mathbb{C}^K$, with $\norm{\bvphi_k}^2 \!= \!1$.
The pilots are thus drawn from a set of $K$ orthonormal sequences.
Denoting by $\etap_k$ the UL transmit power of the UE $k$ during the training phase, assumed to be constant over the $\Nr + 1$ intervals, the pilot signal received at the BS during the interval $t$, $t\!=\!1,\ldots,\Nr+1$, when the RIS configuration is characterized by $\bP_t\!=\!\diag(\bp_t)$, is given by the $(\Na\! \times \! K)$-dimensional matrix  
\begin{align}
\label{eq:Yp}	
\bYp_t 
&= \sum\limits_{k=1}^K \sqrt{K\etap_k} \bG_k \bp_t \bvphi_k\trans \!+\! \delta\bH \bP_t \wtbNp_t \!+\! \bNp_t
\end{align}
where $\bNp_t$ is additive i.i.d. complex Gaussian \textit{static} noise at the BS, while the term $\delta\bH\bP_t\wtbNp_t$ represents a \textit{dynamic} noise contribution occurring in the case of active RIS, with the entries of $\wtbNp_t$ being i.i.d. complex Gaussian RVs, and $\delta$ being a binary coefficient used to distinguish the case of active RIS ($\delta = 1$) from the case of passive RIS ($\delta = 0$).
The columns of $\bG_k$, i.e., ${[\bG_k]}_{:,r}~\forall r$, correspond to the cascaded channels from UE $k$ to the BS through each element $r$ of the RIS. Then,~\eqref{eq:Yp} can be rewritten as
\begin{equation}
\label{eq:Yp:G}
\bYp_t = \! \sum_{k=1}^K \sqrt{K\etap_k} \sum^{\Nr}_{r=1} p_{r,t} {[\bG_k]}_{:,r}  \bvphi_k\trans \!+\! \delta\bH \bP_t \wtbNp_t \!+\! \bNp_t \, ,
\end{equation} 
where $p_{r,t} \in \mathbb{C}$ is the phase-shift introduced by element $r=1,\ldots,\Nr$, in the training interval $t$. Notice that ${[\bG_k]}_{:,r}$ is affected by the same phase-shift $p_{r,t}$ in the $t$-th training interval. Let $\bp_r = [p_{r,0}~p_{r,1}~\cdots~p_{r,\Nr}]\trans$, in order to estimate an arbitrary ${[\bG_k]}_{:,r}$, $\bP_t$ is designed so that 
$\bp\herm_r \bp_q = 0,~ r\neq q$.
Firstly, $\bYp_t$ is projected along the UE $k$ pilot sequence as $\byp_{k,t} = \! \bYp_t \bvphi^{\ast}_k$, which gives an $\Na$-dimensional vector
\begin{align}
	\label{eq:yp_k}
	\byp_{k,t} &= \sqrt{K\etap_k} \sum\nolimits^{\Nr}_{r=1} p_{r,t} {[\bG_k]}_{:,r} \!+\! \delta\bH \bP_t \wtbnp_{k,t} \!+\! \bnp_{k,t} \, ,
\end{align}	 
and $\wtbnp_{k,t} \!=\! \wtbNp \bvphi^{\ast}_k \!\sim\! \CG{\bzero_{\Na}}{\sigma^2_\tR\bI_{\Na}}$, while $\bnp_{k,t} \!=\! \bNp\bvphi^{\ast}_k \!\sim\! \CG{\bzero_{\Na}}{\sigma^2_\tA\bI_{\Na}}$.
Then, $\byp_{k}$ is processed as
\begin{align}
\!\bzp_{k,r} \!&= \!\frac{1}{\sqrt{\Nr\!+\!1}} \sum\limits^{\Nr}_{t=0} {p^{\ast}_{r,t}} \by^{\bp}_{k,t} \nonumber \\
&= \!\sqrt{\frac{K\etap_k}{\Nr\!+\!1}} \!\sum^{\Nr}_{t=0} |p_{r,t}|^2 {[\bG_k]}_{:,r} \!+\! \wtbnp_{k} + \bnp_{k}\nonumber \\
&= \!\sqrt{K\etap_k(\Nr\!+\!1)} {[\bG_k]}_{:,r} + \wtbnp_{k} + \bnp_{k} \, ,   
\label{eq:sufficient-statistic}
\end{align}
where
\begin{align}
\bnp_{k} &= \sqrt{\frac{1}{\Nr\!+\!1}} \sum^{\Nr}_{t=0} p^{\ast}_{r,t} \bnp_{k,t} \sim \CG{\bzero}{\sigma^2_\tA \bI_{\Na}} \, , \\
\wtbnp_{k} &= \frac{\delta}{\sqrt{\Nr\!+\!1}} \bH\! \sum^{\Nr}_{t=0} \! |p_{r,t}|^2  \wtbnp_{k,t} \sim \CG{\bzero}{\delta \bH\bH\herm \sigma^2_\tR} \, .
\end{align}
The \textit{sufficient statistic} in~\eqref{eq:sufficient-statistic} serves to compute the minimum mean square error (MMSE) estimate ${[\widehat{\bG}_k]}_{:,r} \!=\! \bR_{k,r} \boldsymbol{\Psi}_{k,r}^{-1} \bzp_{k,r}$,
where 
\begin{equation}
\bR_{k,r} = \sqrt{K\etap_k(\Nr\!+\!1)} \sum\limits^{\Nr}_{q=1}[\bR_k]_{r,q} {[\bH]}_{:,r} {[\bH]}\herm_{:,q} \, ,
\end{equation}
and $\boldsymbol{\Psi}_{k,r} \!=\! \EX{\bzp_{k,r} (\bzp_{k,r})\herm}$ is the correlation matrix of the signal in~\eqref{eq:sufficient-statistic}, which is given in closed form by
\begin{align}
\label{eq:MMSE:Psi}
\boldsymbol{\Psi}_{k,r} \!=\! \sqrt{\etap_k K(\Nr\!+\!1)} \bR_{k,r} \!+\! \delta \bH\bH\herm \sigma^2_\tR + \sigma^2_\tA \bI_{\Na} \, .
\end{align}%

\section{DL Performance {Measure and Problem Formulation}}
\label{sec:performance-formulation}
During the DL data transmission phase, the signal received by the UE $k$ in the $\tau$-th channel use (we omit the dependency on $\tau$ for brevity) is given by
\begin{align}
r_k = &\sqrt{\etad_k} \bp\trans \bG\trans_k \bw_k x_k \!+\!\! \sum_{j\neq k}^K \!\sqrt{\etad_j} \bp\trans \bG\trans_k \bw_j x_j \!+\! \delta\wt{z}_k \!+\! z_k,
\label{eq:downlink_data}
\end{align}
where $\etad_k$ is the DL transmit power, $\bw_k \!\in\!\mC^{\Na}$ denotes the precoding vector, and $x_k$ is the data symbol intended for the UE $k$, with $\EX{|x_k|^2}\!=\!1$. The term $z_k \!\sim\!\CG{0}{\sigma^2_k}$ is additive white Gaussian noise at the UE $k$, while $\delta\wt{z}_k$ denotes the dynamic noise introduced by the active RIS in the DL, with $\wt{z}_k \!=\! \bh_k\trans \bP \bz_\tR$, and $\bz_\tR \!\sim\! \mathcal{CN}(\bzero_{N_\tR}, \sigma^2_{\tR} \bI_{\Nr})$.
The precoding vector can be obtained via the \textit{UL-DL duality theorem}~\cite[Section 4.3.2]{massivemimobook} as
$\bw^{\text{x}}_{k,\text{y}} \!=\! \big(\bv^{\text{x}}_{k,\,\text{y}}\big)^{\!\ast}/\big\Vert\bv^{\text{x}}_{k,\,\text{y}}\big\Vert$, 
where, $\bv^{\text{x}}_{k,\,\text{y}}$ is the combining vector used for the UL data detection, $\text{x}\!=\!\{\text{pCSI},\text{iCSI\}}$ indicates whether perfect CSI (pCSI) or imperfect CSI (iCSI) is available at the RIBS, and $\text{y} \!=\! \{\text{MMSE},\,\text{MR}\}$ denotes the considered precoding scheme, i.e., either MMSE or maximum ratio (MR).
The MMSE combining vector under pCSI assumption is given by~\cite[Appendix C.3.2]{massivemimobook} $\bv\pCSI_{k,\text{MMSE}} \!=\! \eta\up_k \boldsymbol{\Upsilon}\inv \bG_k\bp \,$, 
where $\boldsymbol{\Upsilon} \!=\! \sum\nolimits^K_{i=1} \eta\up_i \bG_i\bp\bp\herm\bG\herm_i \!+\! \delta \sigma^2_\tR  \bH \bP \bP\herm \bH\herm \!+\! \sigma^2_\tA \bI_{\Na}$ and $\eta\up_i$ is the UL data transmit power of UE $i$. While, the MR combining vector is $\bv\pCSI_{k,\text{MR}} \!=\! \bG_k\bp$.

\subsection{{Spectral Efficiency}}
Under the assumption of iCSI knowledge at the RIBS, the combining vectors are designed upon the channel estimates, namely $\bv\iCSI_{k,\text{y}} \!=\! \bv\pCSI_{k,\text{y}}\big|_{\bG_k = \wh{\bG}_k}$.

Given~\eqref{eq:downlink_data}, the DL \textit{signal-to-interference-plus-noise} ratio (SINR) at the UE $k$, for any precoding scheme, under the assumption of pCSI knowledge at the RIBS and at the UEs is
\begin{equation}
	\gamma\pCSI_{\text{d},k} (\bp,\,\bEtad,\bG) \!=\! \frac{\etad_k \left| \bp\trans \bG\trans_k \bw\pCSI_k\right|^2}{\sum\limits_{j\neq k}^K \etad_j\left|\bp\trans	\bG_k\trans \bw\pCSI_j \right|^2 \!+\! \delta\norm{\bp\trans \bH_k}^2\! \sigma^2_\tR \!+\! \sigma^2_k} \, ,
	\label{eq:DL_SINR_PCSI}
\end{equation}
where $\bEtad \!=\! [\etad_1~\cdots~\etad_K]\trans$.
Consequently, assuming infinite-length codewords, an achievable ergodic spectral efficiency (SE), in bit/s/Hz, for the UE $k$ is obtained by using the classical Shannon expression
\begin{equation}
	\text{SE}\pCSI_{\text{d},k} = \xi \EX{\log_2 \Big(1 + \gamma\pCSI_{\text{d},k}(\bp,\,\bEtad,\bG)\Big)},
	\label{eq:DL_SE_PCSI}
\end{equation}
where $0\!<\!\xi\!<\!1$ accounts for the fraction of the coherence block used for the DL data transmission, and the expectation is taken with respect to the small-scale fading quantities.

By assuming iCSI knowledge at the BS and genie-aided UEs that know the instantaneous channel gain, an upper bound on the DL achievable SE can be obtained by~\cite{Caire_Bounds2018}
\begin{equation}
	\text{SE}^{\text{g-aided}}_{\text{d},k} = \bar{\xi}~\EX{\log_2 \Big(1+ \gamma\iCSI_{\text{d},k}(\bp,\,\bEtad,\bG)\Big)}\, ,
	\label{eq:DL_SE_GAIDED}
\end{equation}
where the prelog factor $0\!<\!\bar{\xi}\!<\!1$ accounts for the fraction of the coherence interval used for DL data transmission, and for the pilot overhead as $\bar{\xi} = \xi [1-K(\Nr+1)/\tc]$.
In~\eqref{eq:DL_SE_GAIDED}, the SINR value $\gamma\iCSI_{\text{d},k}$ is obtained from the r.h.s. of~\eqref{eq:DL_SINR_PCSI} but replacing $\{\bw\pCSI_k\}$ with $\{\bw\iCSI_k\}$, and the expectation is taken with respect to the small-scale fading quantities.


\subsection{{Minimum-SE Maximization Problem}}

The SE depends on the RIS configuration, through the vector $\bp$, on the DL power control coefficients $\bEtad$ and on the precoding vectors $\{\bw_k\}$. In this paper, in order to show the potentialities of the proposed RIBS architecture, we consider the joint optimization of the RIS configuration and of the transmit power coefficients to maximize the minimum per-UE SE throughout all the network, under power budget and hardware constraints. 
The optimization problem formulates as%
\begin{subequations} \label{prob:P0}
\begin{align}	
  \mathop {\text{maximize}}\limits_{\bp,\,\bEtad,\, \varepsilon} & \quad \min\limits_k \gamma_{\text{d},k}^{\text{x}}(\bp,\bEtad, \bar{\bG})  \label{prob:P0:obj} \\
  \text{s.t.}%
    &\quad \sum\nolimits^K_{k=1} \etad_k \!\leq\! (1\!-\!\varepsilon)P^{\text{dl}}_{\text{tx,max}} \, , \label{prob:P0:C1} \\    
    &\quad P^{\RIS}_{\text{tx}}(\bp,\bEtad) \leq \varepsilon P^{\text{dl}}_{\text{tx,max}} \, , \label{prob:P0:C2} \\
    &\quad \big|[\,\bp\,]_i\big| \!\leq\! \amax	,~\forall i=1,\ldots, \Nr \, ,\label{prob:P0:C3} \\[-0.5ex]
    &\quad \etad_k \geq 0, \; \forall\,k \!=\! 1,\ldots,K \, , \label{prob:P0:C4} \\[-0.5ex]
    &\quad 0\!<\varepsilon\!<1 \, , \label{prob:P0:C5}
\end{align}
\end{subequations}
where $\text{x}\!=\!\{\text{pCSI},\,\text{iCSI\}}$ and $\bar{\bG} \!=\! \{\bG, \, \wh{\bG}\}$, according to the information available at the BS. The available overall DL transmit power budget is denoted by $P^{\text{dl}}_{\text{tx,max}}$, and $\amax$ denotes the maximum amplification factor of the single RIS element.
In the case of active RIBS, the DL power budget should be optimally split between the RIS and the BS. Hence, the optimization is also carried out with respect to $\varepsilon$ defined as the fraction of $P^{\text{dl}}_{\text{tx,max}}$ allocated to the RIS and $(1-\varepsilon)P^{\text{dl}}_{\text{tx,max}}$ represents the transmit power available at the BS. While, $P^{\RIS}_{\text{tx}}(\bp,\bEtad)$ denotes the effective transmit power at the RIS, which is function of both $\bp$ and $\bEtad$.  
In the case of passive RIBS, $P^{\text{dl}}_{\text{tx,max}}$ is entirely allocated at the BS, thus $\varepsilon\!=\!0$ and the constraints~\eqref{prob:P0:C2},~\eqref{prob:P0:C5} are deactivated. Moreover, it holds that the entries of $\bp$ must have unit modulus, i.e., $\amax\!=\!1$.

Ideally, the optimization of the RIS phase-shifts and of the DL transmit powers should be joint. However, such a joint optimization leads to a well-known non-convex problem with respect to $\bp$ and $\bEtad$. {To circumvent this limitation, in the next section we propose a practical solution to Problem \eqref{prob:P0}, for both active and passive RIS configurations.}

\section{{System Optimization}} 
\label{sec:optimization}
We resort to the alternating optimization approach~\cite{BertsekasNonLinear} to solve Problem \eqref{prob:P0}. We firstly initialize the values of $\bEtad$ and solve the problem with respect to $\bp$. Then, given the optimal values of $\bp$, we solve the problem with respect to $\bEtad$. The process repeats until the objective function converges or the maximum number of iterations is reached.

\subsection{Optimization of the RIS Phase-shifts}
\label{subsec:RIS_optimization}
While in a traditional communication system UEs' channels are fully determined by the propagation environment, in a RIS-aided communication system the propagation environment can be conveniently shaped by fine-tuning the elements of the RIS. The BS is in charge of the RIS fine-tuning, which is carried out upon the acquired CSI to maximize the minimum per-UE SE throughout all the network. If perfect CSI knowledge is assumed, then the BS is able to exactly compute~\eqref{eq:DL_SE_PCSI} upon the information available and to optimize the RIS configuration accordingly. 
If imperfect CSI knowledge is assumed, then the BS is not able to exactly compute~\eqref{eq:DL_SE_GAIDED}, as $\bG$ is not available. In this regard, we assume that the BS cooks up its own SE metric upon the channel estimates as
\begin{align}
	\text{SE}^{\text{iCSI}}_{\text{d},k} &= \bar{\xi}~\EX{\log_2 \Big(1+ \gamma\iCSI_{\text{d},k}(\bp,\,\bEtad,\wh{\bG})\Big)} \, ,
	\label{eq:DL_SE_ICSI}
\end{align}

\vspace{-5mm}
\begin{align}
	\gamma\iCSI_{\text{d},k}(\bp,\,\bEtad,\wh{\bG}) &\!=\! \frac{\etad_k \left| \bp\trans \wh{\bG}\trans_k \bw\iCSI_k\right|^2}{\sum\limits_{j\neq k}^K \etad_j\left|\bp\trans	\wh{\bG}_k\trans \bw\iCSI_j \right|^2 \!+\! \delta\norm{\bp\trans \wh{\bH}_k}^2\! \sigma^2_\tR \!+\! \sigma^2_k} \, , 
\end{align}
where $\wh{\bH}_k\!=\!\diag(\wh{\bh}_k)$ and $\wh{\bh}_k$ is the \textit{least-squares} solution of the system of equations $[\,\wh{\bh}_k\,]_r [\,\bH\,]_{:,r} \!=\! [\,\wh{\bG}_k\,]_{:,r}$. 
The SE expression in~\eqref{eq:DL_SE_ICSI} is used by the BS only as a support to the RIS optimization in case of iCSI knowledge.

\subsubsection{Passive RIS} In this case, the entries of $\bp$ must have unit modulus (i.e., hardware constraint), and $\delta \!=\! 0$. 
Hence, the optimization problem is formulated as
\begin{subequations} \label{prob:P1:P:passive}
\begin{align}	
  \mathop {\text{maximize}}\limits_{\bp} & \quad \min\limits_k \gamma_{\text{d},k}^{\text{x}}(\bp, \bar{\bG})  \label{prob:P1:obj} \\
  \text{s.t.} &\quad \big|[\,\bp\,]_i\big| = 1,~\forall i=1,\ldots, \Nr \, . \label{prob:P1:C1}
\end{align}
\end{subequations}
The power coefficients $\bEtad$ are constant and pre-determined at this stage.
Problem~\eqref{prob:P1:P:passive} is nonconvex, due to both the nonconvexity of the objective function and of the constraints. 
Firstly, we relax the constraint by setting $\big|[\,\bp\,]_i\big| \!\leq\! 1, \, \forall i \!=\!1,\ldots, \Nr$. The optimal solution will be finally normalized so as to meet the element-wise unit modulus constraint. Then, we rewrite~\eqref{prob:P1:P:passive} in epigraph form~as%
\begin{subequations} \label{prob:P2:P:passive}
\begin{align}	
  \mathop {\text{maximize}}\limits_{\bp, \, \varrho} & \quad \varrho \label{prob:P2:obj} \\[-1ex]
  \text{s.t.} 
  	&\quad \gamma_{\text{d},k}^{\text{x}}(\bp,\bar{\bG}) \geq \varrho \, ,\label{prob:P2:C1} \\
    &\quad \big|[\,\bp\,]_i\big| \!\leq\! 1,~\forall i=1,\ldots, \Nr \, .\label{prob:P2:C2}
\end{align}
\end{subequations}
Problem~\eqref{prob:P2:P:passive} can be efficiently solved with respect to $\varrho$ via \textit{bisection method}~\cite{Boyd2004}, but it is still nonconvex in $\bp$.
Next, we convexify the constraint~\eqref{prob:P2:C1}. Let $\text{num}(\bp,\bar{\bG})$ and $\text{den}(\bp,\bar{\bG})$ be the numerator and the denominator of the effective SINR, $\gamma_{\text{d},k}^{\text{x}}(\bp,\bar{\bG})$, respectively. Then,~\eqref{prob:P2:C1} can be written as
\begin{equation}
\label{prob:P2:C1:difference-of-convex}
\varrho~\text{den}(\bp,\bar{\bG}) - \text{num}(\bp,\bar{\bG}) \leq 0 \, .
\end{equation}
Let us enforce the precoding vectors to be constant with respect to $\bp$. Under these circumstances, both $\text{num}(\bp,\bar{\bG})$ and $\text{den}(\bp,\bar{\bG})$ are quadratic functions, hence convex functions, in $\bp$. However, even for a fixed value of $\varrho$ (i.e., for each iteration of the bisection search),~\eqref{prob:P2:C1:difference-of-convex} is still nonconvex since it is the difference of two convex functions.
As any convex function is lower-bounded by its Taylor expansion around any given point, $\bp_0$, a convex constraint is obtained by replacing $\text{num}(\bp,\bar{\bG})$ with its first-order Taylor expansion $\text{num}(\bp,\bp_0,\bar{\bG})$ in~\eqref{prob:P2:C1:difference-of-convex}~as  
\begin{equation}
\label{prob:P2:C1:convexification}
\varrho~\text{den}(\bp,\bar{\bG}) - \text{num}(\bp,\bp_0,\bar{\bG}) \leq 0 \, ,
\end{equation}
where $\text{num}(\bp,\bp_0,\bar{\bG}) \!=\! \bp_0\trans \bA_{k,k} \bp^{\ast}_0 \!+\! 2\real{\bp_0\herm \bA_{k,k} (\bp \!-\! \bp_0)}$, and the $(\Nr\!\times\!\Nr)$-dimensional matrix $\bA_{k,j}$ is given, according to the information available at the BS, by
\begin{equation}
\label{eq:Akj}
\bA_{k,j} = \begin{cases}
	\etad_k \bG_k\trans \bw_k\pCSI \big(\bw_j\pCSI\big)\herm \bG^{\ast}_k\, , \qquad &\text{for pCSI,} \\
	\etad_k \wh{\bG}_k\trans \bw_k\iCSI \big(\bw_j\iCSI\big)\herm \wh{\bG}^{\ast}_k\, , \qquad &\text{for iCSI,}
\end{cases}
\end{equation}
which is symmetric and positive semidefinite. 
Hence, at the $n$-th iteration of the bisection search over $\varrho$, we solve the following feasibility problem by using on-the-shelf optimization tools (e.g., CVX~\cite{cvx2014}):
\begin{subequations} \label{prob:P3:P:passive}
\begin{align}	
  \mathop {\text{find}} & \quad \bp_n \label{prob:P3:obj} \\
  \text{s.t.} 
  	&\quad \varrho~\text{den}(\bp_n,\bar{\bG}) - \text{num}(\bp_n,\bp_{n-1},\bar{\bG}) \leq 0 \, , \label{prob:P3:C1} \\
    &\quad \big|[\,\bp_n\,]_i\big| \!\leq\! 1,~\forall i=1,\ldots, \Nr \, .\label{prob:P3:C2}
\end{align}
\end{subequations}
where 
\begin{align}
&\text{den}(\bp_n,\bar{\bG}) \!=\! \sum\limits_{j\neq k}^K \bp_n\trans \bA^{(n-1)}_{k,j} \bp_n^{\ast} \!+\! \delta \, \bp_n\trans \wt{\bA}_k \bp_n^{\ast} \!+\! \sigma^2_k \, , \label{eq:den_pn}\\
&\text{num}(\bp_n,\bp_{n-1},\bar{\bG}) \!=\! \bp_{n-1}\trans \bA^{(n-1)}_{k,k} \bp^{\ast}_{n-1} \nonumber \\
&\hspace*{3cm}\!+\! 2\real{\bp_{n-1}\herm \bA^{(n-1)}_{k,k} (\bp \!-\! \bp_{n-1})} , \label{eq:num_pn} 
\end{align} 
and $\bA^{(n-1)}_{k,j}$ is computed as in~\eqref{eq:Akj}, but with the precoding vectors designed upon the phase-shift vector at the previous iteration, $\bp_{n-1}$, while 
$\wt{\bA}_k \!=\! \bH_k \bH^{\ast}_k$ in case of pCSI or $\wt{\bA}_k \!=\! \wh{\bH}_k \wh{\bH}^{\ast}_k$ in case of iCSI.
Algorithm~\ref{alg:RIS_opt} summarizes the steps taken to optimize the RIS phase-shits. %
\setlength{\textfloatsep}{5mm}
\begin{algorithm}[t]\small
	\setstretch{1}
	\caption{Optimization of the passive RIS phase-shifts}
	\vspace{1mm}
	\textbf{Input:} A feasible $\bp_0$, $n\!=\!1$, tolerance $\epsilon\!>\!0$, $\bEtad$, CSI to compute $\gamma_{\text{d},k}^{\text{x}}(\bp_0,\bar{\bG})$, $\varrho_{\text{max}}$ and $\varrho_{\text{min}}$ selected from the range of relevant values of $\gamma_{\text{d},k}^{\text{x}}(\bp_0,\bar{\bG})$;
	\begin{algorithmic}[1]
		\State Compute $\{\bw^{\text{x}}_{k}\}$ upon $\bp_0$ and $\{\bA^{(0)}_{k,j}\}$ accordingly;
		\Repeat {\texttt{~\%\%  Bisection algorithm}}		
		\State  $\varrho \gets (\varrho_{\text{min}} + \varrho_{\text{max}})/2$;
		\State Solve feasibility problem~\eqref{prob:P3:P:passive};
		\If{\eqref{prob:P3:P:passive} is unfeasible}{~$\varrho_{\text{max}} \gets \varrho$;}
		\Else		
		\State $\varrho_{\text{min}} \gets \varrho$;
		\State $\bp^{\star} \gets \bp_n$;
		\State Compute $\{\bw^{\text{x}}_{k}\}$ upon $\bp_n$ and $\{\bA^{(n)}_{k,j}\}$ accordingly;
		\State $n \gets n\!+\!1$;
		\EndIf
		\Until $\varrho_{\text{max}}\!-\!\varrho_{\text{min}} \leq \epsilon$
	\end{algorithmic}
	\textbf{Output:} $\bp \!=\! \exp(\,\imagj\, \arg(\bp^{\star})\,)$.
	\label{alg:RIS_opt}
\end{algorithm}
Its initial value, $\bp_0$, can be selected at random and eventually normalized to ensure unitary modulus as
\begin{equation}
\label{eq:p-scaling:passive}
\bp_0 \gets \exp(\,\imagj\, \arg(\bp_0)\,) \, .
\end{equation}%
At each iteration of the bisection search over $\varrho$, whenever the feasibility problem~\eqref{prob:P3:P:passive} is solved, the precoding vectors and the matrices $\{\bA^{(n)}_{k,j}\}$ must be updated accordingly. The output of Algorithm~\ref{alg:RIS_opt} is the RIS phase-shift vector, $\bp$, normalized so that its entries have unit modulus. {Manifold optimization~\cite{junior2024}, majorization-minimization and semidefinite relaxation~\cite{Zhou2024} are effective alternative solutions widely used for optimization problems involving unit-modulus constraints, like Problem~\eqref{prob:P1:P:passive}. These optimization methods could potentially outperform Algorithm~\ref{alg:RIS_opt}. However, their direct comparison with the proposed algorithm is out of the scope of this paper.}
\subsubsection{Active RIS} In this case, the entries of $\bp$ are constrained to have modulus smaller than a maximum value, denoted by $\amax$, and $\delta \!=\! 1$.
In addition, the active RIS is subject to a power constraint that is function of $\bp$.
In line with the power consumption model presented in~\cite{Long2021}, we 
define $P_r$ as the power consumed by the $r$-th reflecting element of the active RIS with $r \!=\! 1,\ldots,\Nr$, given by the sum of a traffic-independent component (TIP) and a traffic-dependent component (TDP). The former includes the switch and control circuit power consumption, $P_{\tC}$,  and the digital converter biasing power consumption, $P_{\text{dc}}$. 
The latter is a linear function of the active RIS transmit power $P_{\text{tx},r} (\bp,\bEtad)$.
Hence, the total power consumption at the active RIS, assuming $\Nr$ identical reflecting elements, is given by
\begin{equation}
\label{eq:power-budget-active-RIS}
P_{\RIS} \!=\! \sum\limits^{\Nr}_{r=1} P_{r} \!=\! \Nr(P_{\tC} \!+\! P_{\text{dc}}) + \xi \sum\limits^{\Nr}_{r=1} \!P_{\text{tx},r} (\bp,\bEtad)   \, ,
\end{equation}
where $\xi$ is the efficiency of the amplifiers. $P_{\text{tx},r}(\bp,\bEtad)$ is the only term of~\eqref{eq:power-budget-active-RIS} being dependent on $\bp$ and $\bEtad$, and thereby involved in the optimization. We can omit the dependence on $\bEtad$ at this stage, and compute $P^{\RIS}_{\text{tx}} (\bp)$. In the DL, the received signal at the active RIS, coming from the BS, is
\begin{equation}
\br_\RIS \!=\! \sum\nolimits^K_{j=1}\sqrt{\etad_j}\bH\trans \bw_j x_j + \bz_\tR \, .
\end{equation}
The active RIS shifts $\br_\RIS$ by applying $\bP$ and retransmits 
\begin{equation}
\bar{\br}_\RIS \!=\! \bP\br_\RIS \!=\! \sum\nolimits_{j=1}^K \sqrt{\etad_j} \bP \bH\trans \bw_j x_j \!+\! \bP \bz_\tR \, ,
\end{equation}
whose transmit power is given by
\begin{align}
\label{eq:RIS-tx-power}
P^{\RIS}_{\text{tx}} (\bp) &\!=\! \sum\nolimits^{\Nr}_{r=1} P_{\text{tx},r}(\bp) \!=\! \EX{\norm{\bar{\br}_\RIS}^2} \!=\! \tr\left(\EX{\bar{\br}_\RIS\bar{\br}\herm_\RIS}\right) \nonumber \\
&\!=\! \sum\nolimits^K_{j=1} \etad_j \tr( \bP \bH\trans \bw_j \bw\herm_j \bH^{\ast} \bP\herm) \!+\! \sigma^2_\tR \tr(\bP\bP\herm) \nonumber \\
&\!=\! \tr\left(\bP \bH\trans \bar{\bW} \bH^{\ast} \bP\herm  \right) \!+\! \sigma^2_\tR \tr(\bP\bP\herm) \nonumber \\
&\!=\! \tr\left(\bP \bB \bP\herm  \right) \stackrel{(a)}{=} \bp\herm \wt{\bB} \bp \, .
\end{align}
where 
\begin{align}
    \bar{\bW} \!=\! \sum\nolimits^K_{j=1} \etad_j \bw_j \bw\herm_j \, ,
\end{align}
$\bB \!=\! \bH\trans \bar{\bW} \bH^{\ast} \!+ \sigma^2_\tR \bI_{\Nr}$, and $\wt{\bB}$ is the diagonal matrix built upon the diagonal entries of $\bB$, namely $\wt{\bB} \!=\!\diag(\diag(\bB))$. Indeed, $(a)$ follows the identity $\tr(\bP\bB\bP\herm) \!=\! \sum_i [\bP]_{i,i} [\bB]_{i,i} [\bP]^{\ast}_{i,i}$, for $\bP$ being diagonal.
The minimum SE maximization problem can be solved by using the same methodology described for the passive RIS, but with a slightly different formulation of the feasibility problem:
\begin{subequations} \label{prob:P1:P:active}
\begin{align}	
  \mathop {\text{find}} & \quad \bp_n \label{prob:P1:P:active:obj} \\
  \text{s.t.} 
  	&\quad \varrho~\text{den}(\bp_n,\bar{\bG}) - \text{num}(\bp_n,\bp_{n-1},\bar{\bG}) \leq 0 \, , \label{prob:P1:P:active:C1} \\
    &\quad \big|[\,\bp_n\,]_i\big| \!\leq\! \amax	,~\forall i=1,\ldots, \Nr \, ,\label{prob:P1:P:active:C2} \\
    &\quad \bp\herm \wt{\bB} \bp \leq \varepsilon P^{\text{dl}}_{\text{tx,max}} \, . \label{prob:P1:P:active:C3} 
\end{align}
\end{subequations}
Constraint~\eqref{prob:P1:P:active:C3} is convex in $\bp$, hence problem~\eqref{prob:P1:P:active} can be efficiently solved.
Algorithm~\ref{alg:RIS_opt_active} summarizes the steps taken to optimize the active RIS phase-shits. %
\setlength{\textfloatsep}{5mm}
\begin{algorithm}[t] \small
	\setstretch{1}
	\caption{Optimization of the active RIS phase-shifts}
	\vspace{1mm}
	\textbf{Input:} A feasible $\bp_0$, $n\!=\!1$, tolerance $\epsilon\!>\!0$, $\bEtad$, CSI to compute $\gamma_{\text{d},k}^{\text{x}}(\bp_0,\bar{\bG})$, $\varrho_{\text{max}}$ and $\varrho_{\text{min}}$ selected from the range of relevant values of $\gamma_{\text{d},k}^{\text{x}}(\bp_0,\bar{\bG})$, $\amax$, $\varepsilon P^{\text{dl}}_{\text{tx,max}}$.
	\begin{algorithmic}[1]
		\State Compute $\{\bw^{\text{x}}_{k}\}$ upon $\bp_0$ and $\{\bA^{(0)}_{k,j}\}$ accordingly;
		\Repeat {\texttt{~\%\%  Bisection algorithm}}		
		\State  $\varrho \gets (\varrho_{\text{min}} + \varrho_{\text{max}})/2$;
		\State Solve feasibility problem~\eqref{prob:P1:P:active};
		\If{\eqref{prob:P1:P:active} is unfeasible}{~$\varrho_{\text{max}} \gets \varrho$;}
		\Else		
		\State $\varrho_{\text{min}} \gets \varrho$;
		\State $\bp^{\star} \gets \bp_n$;
		\State Compute $\{\bw^{\text{x}}_{k}\}$ upon $\bp_n$ and $\{\bA^{(n)}_{k,j}\}$ accordingly;
		\State $n \gets n\!+\!1$;
		\EndIf
		\Until $\varrho_{\text{max}}\!-\!\varrho_{\text{min}} \leq \epsilon$
	\end{algorithmic}
	\textbf{Output:} $\bp^{\star}$.
	\label{alg:RIS_opt_active}
\end{algorithm}
An initial set of feasible phase-shifts for the active RIS, $\bp_0$, can be selected at random and eventually normalized to meet the constraints~\eqrefs{prob:P1:P:active:C2}{prob:P1:P:active:C3} as follows: let $\wt{\bp}_0$ be a vector of randomly-drawn complex-valued elements. Firstly, we scale the modulus of those elements of $\wt{\bp}_0$ exceeding $\amax$ as
\begin{align}
\label{eq:p-scaling:active}
[\,\wt{\bp}_0\,]_i & \gets \amax\exp(\imagj \arg(\,[\,\wt{\bp}_0\,]_i\,), \quad \forall\,i\!\in\!\mathcal{I},
\end{align}
where $\mathcal{I} \!=\!\{i:\left|[\,\wt{\bp}_0\,]_i\right| \!>\! \amax\}$.
We obtain $\bp_0$ by scaling $\wt{\bp}_0$ as
\begin{equation}
\label{eq:p-adjusting:active}
\bp_0 \! = \! \sqrt{\frac{\varepsilon P^{\text{dl}}_{\text{tx,max}}}{P^{\RIS}_{\text{tx}}\left(\wt{\bp}_0\right)}}\, \wt{\bp}_0 \, ,
\end{equation}  
which guarantees that the transmit power constraint~\eqref{prob:P1:P:active:C3} is met for a pre-determined set of power coefficients $\bEtad$. As an alternative to~\eqref{eq:p-adjusting:active}, we can simply select $\bp_0 \! = \! \wt{\bp}_0$, with $\wt{\bp}_0$ given by~\eqref{eq:p-scaling:active}, and scale the power coefficients $\bEtad$ as
\begin{equation}
\label{eq:eta-adjusting:active}
\bEtad \!\gets\! \frac{\varepsilon P^{\text{dl}}_{\text{tx,max}}}{P^{\RIS}_{\text{tx}}\left(\wt{\bp}_0\right)} \bEtad \;.
\end{equation}%

\subsection{Optimization of the DL Transmit Power}
\label{subsec:power-control}
We now consider the problem of optimizing the DL transmit powers, $\bEtad$, while the RIS phase-shifts are pre-determined.%
\subsubsection{Passive RIS} In this case $\delta \!=\! 0$, hence the optimization problem is formulated as
\begin{subequations} \label{prob:P1:eta:passive}
\begin{align}	
  \mathop {\text{maximize}}\limits_{\bEtad} & \quad \min\limits_k \gamma_{\text{d},k}^{\text{x}}(\bEtad, \bar{\bG})  \label{prob:P1:eta:obj} \\[-.5ex]
  \text{s.t.} &\quad \sum\nolimits^K_{k=1} \etad_k \!\leq\! P^{\text{dl}}_{\text{tx,max}} \, , \label{prob:P1:eta:C1} \\
  &\quad \etad_k \geq 0, \; \forall\,k \!=\! 1,\ldots,K \, . \label{prob:P1:eta:C2} 
\end{align}
\end{subequations}
Notice that $\bp$ is a constant and pre-determined at this stage.
Problem~\eqref{prob:P1:eta:passive} can be written in epigraph form as
\begin{subequations} \label{prob:P2:eta:passive}
\begin{align}	
  \mathop {\text{maximize}}\limits_{\bEtad, \, \varrho} & \quad \varrho \label{prob:P2:eta:obj} \\[-1.5ex]
  \text{s.t.} 
  	&\quad \frac{\etad_k \bar{g}_{k,k}}{\sum\nolimits_{j\neq k}^K \etad_j \bar{g}_{k,j} \!+\! \delta \bar{h}_k \!+\! \sigma^2_k} \geq \varrho \, ,\label{prob:P2:eta:C1} \\
    &\quad \eqref{prob:P1:eta:C1}\text{\,--\,}\eqref{prob:P1:eta:C2} \nonumber
\end{align}
\end{subequations}
where $\bar{g}_{k,j} \!=\! |\bp\trans\bar{\bG}_k\trans \bw^{\text{x}}_j |^2$, $\bar{h}_k \!=\! \lVert\bp\trans \bar{\bH}_k\rVert^2 \sigma^2_\tR$, and $\bar{\bH}_k \!=\! \{\bH_k, \, \wh{\bH}_k\}$, based on the information available at the BS.
Problem~\eqref{prob:P2:eta:passive} can be solved via bisection search over $\varrho$, in each iteration solving the following convex feasibility problem:
\begin{subequations} \label{prob:P3:eta:passive}
\begin{align}	
  \mathop {\text{find}} & \quad \bEtad \label{prob:P3:eta:passive:obj} \\[-.5ex]
  \text{s.t.} 
    &\quad \etad_k \frac{ \bar{g}_{k,k}}{\varrho} \!-\! \sum\nolimits_{j\neq k}^K \etad_j \bar{g}_{k,j} \!-\! \delta \bar{h}_k \!-\! \sigma^2_k \!\geq\! 0  \, ,\label{prob:P3:eta:passive:C1} \\
    &\quad \eqref{prob:P1:eta:C1}\text{\,--\,}\eqref{prob:P1:eta:C2} \nonumber
\end{align}
\end{subequations}
Algorithm~\ref{alg:power-control} summarizes the procedure detailed above.%
\begin{algorithm}[!t] \small
\caption{Optimal power control with passive RIS}
\textbf{Input: } Pre-determined value of $\bp$, CSI to compute $\gamma_{\text{d},k}^{\text{x}}(\bEtad,\bar{\bG})$, $\varrho_{\text{max}}$ and $\varrho_{\text{min}}$ selected from the range of relevant values of $\gamma_{\text{d},k}^{\text{x}}(\bEtad,\bar{\bG})$, tolerance $\nu \!>\! 0$.
\begin{algorithmic}[1]
		\Repeat {\texttt{~\%\%  Bisection algorithm}}		
		\State  $\varrho \gets (\varrho_{\text{min}} + \varrho_{\text{max}})/2\,;$
		\State Solve feasibility problem~\eqref{prob:P3:eta:passive};
		\IfThenElse{\eqref{prob:P3:eta:passive} is unfeasible}{$\varrho_{\text{max}} \gets \varrho$}{ $\varrho_{\text{min}} \gets \varrho\,;$}
		\Until $\varrho_{\text{max}}\!-\!\varrho_{\text{min}} \leq \nu$
\end{algorithmic}
\textbf{Output: } $\bEtad$.
\label{alg:power-control}
\end{algorithm}%

\subsubsection{Active RIS}
In this case, the available DL transmit power budget, $P^{\text{dl}}_{\text{tx,max}}$, must be properly split between BS and RIS. To this end, a joint optimization of $\varepsilon$ and $\bEtad$ is thus necessary. All this translates into the following feasibility problem:
\begin{subequations} \label{prob:P1:eta:active}
\begin{align}	
  \mathop {\text{find}} & \quad \bEtad, \, \varepsilon \label{prob:P1:eta:active:obj} \\[-.5ex]
  \text{s.t.} 
  	&\quad \etad_k \frac{ \bar{g}_{k,k}}{\varrho} \!-\! \sum\nolimits_{j\neq k}^K \etad_j \bar{g}_{k,j} \!-\! \delta \bar{h}_k \!-\! \sigma^2_k \!\geq\! 0  \, ,\label{prob:P1:eta:active:C1} \\[-.5ex]
    &\quad \sum\nolimits^K_{k=1} \etad_k \!\leq\! (1\!-\!\varepsilon)P^{\text{dl}}_{\text{tx,max}} \, ,\label{prob:P1:eta:active:C2} \\    
    &\quad \bp\herm \wt{\bB} \bp \leq \varepsilon P^{\text{dl}}_{\text{tx,max}} \, , \label{prob:P1:eta:active:C4} \\
    &\quad \etad_k \geq 0, \; \forall\,k \!=\! 1,\ldots,K \, , \label{prob:P1:eta:active:C3} \\
    &\quad 0\!<\varepsilon\!<1 \, , \label{prob:P1:eta:active:C5}
\end{align}
\end{subequations}
where we recall that $\wt{\bB}$ is a linear function of the entries of $\bEtad$, as shown in~\eqref{eq:RIS-tx-power}. Algorithm~\ref{alg:power-control:active} summarizes the steps taken to optimize the power split between the BS and the active RIS, and the transmit powers of the RIBS.%
\begin{algorithm}[!t]\small
\caption{Optimal power control with active RIS}
\textbf{Input: } Pre-determined value of $\bp$, CSI to compute $\gamma_{\text{d},k}^{\text{x}}(\bEtad,\bar{\bG})$, $\varrho_{\text{max}}$ and $\varrho_{\text{min}}$ selected from the range of relevant values of $\gamma_{\text{d},k}^{\text{x}}(\bEtad,\bar{\bG})$, tolerance $\nu \!>\! 0$, $P^{\text{dl}}_{\text{tx,max}}$.
\begin{algorithmic}[1]
		\Repeat {\texttt{~\%\%  Bisection algorithm}}		
		\State  $\varrho \gets (\varrho_{\text{min}} + \varrho_{\text{max}})/2\,;$
		\State Solve feasibility problem~\eqref{prob:P1:eta:active};
		\IfThenElse{\eqref{prob:P1:eta:active} is unfeasible}{$\varrho_{\text{max}} \gets \varrho$}{$\varrho_{\text{min}} \gets \varrho\,;$}
		\Until $\varrho_{\text{max}}\!-\!\varrho_{\text{min}} \leq \nu$
\end{algorithmic}
\textbf{Output: } $\bEtad$, $\varepsilon$.
\label{alg:power-control:active}
\end{algorithm}%
\subsection{RIS Phase-Shifts and DL Powers Alternate Optimization}
\label{subsec:alternating-optimization}
Algorithm~\ref{alg:alternate-optimization} describes the alternate optimization of the RIS phase-shifts and the DL transmit powers.
\begin{algorithm}[!t]\small
\caption{Joint allocation of RIS phase-shifts \& DL powers}
\textbf{Input: } CSI to compute $\gamma_{\text{d},k}^{\text{x}}(\bp, \bEtad,\bar{\bG})$, $P^{\text{dl}}_{\text{tx,max}}$, $\amax$, initial power split $\varepsilon_0$, max number of iterations $\mathcal{I}_{\text{max}}$, $i\!=\!0$, $\chi\!=\! 0.001$.
\begin{algorithmic}[1]
		\Statex {\texttt{\%\%  Initialization}}
		\State 	Draw a random $\bp_0$ and scale it according to~\eqref{eq:p-scaling:passive} or~\eqref{eq:p-scaling:active};
		\State $\bEtad_0 \!\gets\! \boldsymbol{1}_K \cdot [(1\!-\!\varepsilon_0) P^{\text{dl}}_{\text{tx,max}}/K]$;
		\IfThen{$\varepsilon_0\, \texttt{!=}\, 0$}{either adjust $\bp_0$ via~\eqref{eq:p-adjusting:active} or $\bEtad_0$ via~\eqref{eq:eta-adjusting:active}};
		\State Compute $\text{SINR}_0 \gets \min_k \gamma_{\text{d},k}^{\text{x}}(\bp_0, \bEtad_0,\bar{\bG})$;
		\Repeat {\texttt{~\%\%  Alternate Optimization}}		
		\State  $i \gets i+1$;
		\State  Run Alg.~\ref{alg:RIS_opt} or Alg.~\ref{alg:RIS_opt_active}. Let $\bp_i$ denote its output;
		\State Run Alg.~\ref{alg:power-control} or Alg.~\ref{alg:power-control:active}. Let $\bEtad_i$ and $\varepsilon_i$ denote its output;
		\State Compute $\text{SINR}_i \gets \min_k \gamma_{\text{d},k}^{\text{x}}(\bp_i, \bEtad_i,\bar{\bG})$; 
		\Until $\bigg|\dfrac{\text{SINR}_i\!-\!\text{SINR}_{i-1}}{\text{SINR}_{i-1}}\bigg| \!\leq\! \chi$ or $i == \mathcal{I}_{\text{max}}$
\end{algorithmic}
\textbf{Output: } $\bp_i$, $\bEtad_i$, $\varepsilon_i$, $\gamma_{\text{d},k}^{\text{x}}(\bp_i, \bEtad_i, \bar{\bG}),\,\forall\, k \!=\!1,\ldots,K$.
\label{alg:alternate-optimization}
\end{algorithm}%
The initialization step consists in randomly drawing the phase-shifts vector $\bp_0$ and scale its value either according to~\eqref{eq:p-scaling:passive} or~\eqref{eq:p-scaling:active} in case of passive, respectively, active RIS. In addition, the transmit powers $\bEtad_0$ are simply initialized dividing the available power budget at the BS, $(1\!-\!\varepsilon_0) P^{\text{dl}}_{\text{tx,max}}$, by the number of UEs. We notice that $\varepsilon_0 \!=\! 0$ in case of passive RIS. Hence, nonzero values of $\varepsilon_0$ imply that we are in case of active RIS. In this case, the values of either $\bp_0$ or $\bEtad_0$ must be further adjusted via~\eqref{eq:p-adjusting:active} or~\eqref{eq:eta-adjusting:active}, respectively, to meet the power constraint~\eqref{prob:P1:P:active:C3} at the active RIS. The alternate optimization step consists in iteratively running either Algorithm~\ref{alg:RIS_opt} and Algorithm~\ref{alg:power-control} in case of passive RIS or Algorithm~\ref{alg:RIS_opt_active} and Algorithm~\ref{alg:power-control:active} in case of active RIS. The Algorithm for the RIS phase-shifts optimization requires the following input: $\bEtad$, $\amax$ (equal to 1 in case of passive RIS), $P^{\text{dl}}_{\text{tx,max}}$, and $\varepsilon$ (equal to 0 in case of passive RIS). Its output, $\bp$, constitutes the input to the Algorithm for the transmit power optimization. The latter returns the optimal values of $\bEtad$ and $\varepsilon$ (in case of active RIS), which in turn, are the inputs to the Algorithm for the RIS phase-shift optimization at the next iteration of the alternate optimization routine. 
{The convergence of Algorithm~\ref{alg:alternate-optimization} is guaranteed by the alternating optimization framework. In each step, the algorithm fixes one set of variables (either the RIS phase-shifts or the DL transmit powers) and optimizes the other. This ensures that the minimum SINR is non-decreasing at each iteration.  Since the SINR is bounded from below and above (it cannot increase indefinitely), the alternating optimization process will eventually converge to a local optimum, not necessarily the global optimum, yet to a solution where further improvements are not possible. Lastly, Algorithm~\ref{alg:alternate-optimization} uses a stopping criterion based on the convergence of the SINR or reaching a maximum number of iterations. This ensures that the algorithm does not run indefinitely and terminates when the improvement in SINR becomes negligible or when a predefined iteration limit is reached.}%

{\textit{Computational Complexity Analysis:} As for the phase-shift optimization, both Algorithms~\ref{alg:RIS_opt} and~\ref{alg:RIS_opt_active} require to solve quadratically constrained quadratic programs (QCQPs), i.e.,~\eqref{prob:P3:P:passive} and~\eqref{prob:P1:P:active}, respectively. The computational complexity of a QCQP grows polynomially in the number of optimization variables, $N_R$, that is the size of the vector $\bp$. Hence, the complexity of Algorithm~\ref{alg:RIS_opt} and~\ref{alg:RIS_opt_active} coincides and is given by $$\mathcal{O}\left(\log_2\left(\frac{\varrho_{\text{max}}\!-\!\varrho_{\text{min}}} {\epsilon}\right) N_R^{\vartheta}\right),$$ where the logarithmic factor denotes the number of iterations for convergence of the bisection search, while $\vartheta$ is a parameter depending on the implementation of the interior point method (e.g., $\vartheta \!=\! 3.5$~\cite{Mehanna2015}). As for the DL transmit power optimization, Algorithm~\ref{alg:power-control} requires to solve a convex linear problem in $K$ optimization variables, whereas Algorithm~\ref{alg:power-control:active} requires to solve a convex quadratic optimization problem in $K\!+\!1$ optimization variables with a convex quadratic constraint~\eqref{prob:P1:eta:active:C4}. Generalizing, the complexity of the DL power optimization is $$\mathcal{O}\left(\log_2\left(\frac{\varrho_{\text{max}}\!-\!\varrho_{\text{min}}} {\nu}\right) \widetilde{K}\right),$$ where $\widetilde{K}\!=\!K$ for Algorithm~\ref{alg:power-control}, and $\widetilde{K}\!=\!(K\!+\!1)^{\vartheta}$ for Algorithm~\ref{alg:power-control:active}. By combining the above results, the overall complexity of Algorithm~\ref{alg:alternate-optimization} is given by $$\mathcal{O}\!\left(\mathcal{I}_{\text{eff}}\left(\log_2\left(\frac{\varrho_{\text{max}}\!-\!\varrho_{\text{min}}} {\nu}\right)\! \widetilde{K} \!+\! \log_2\left(\frac{\varrho_{\text{max}}\!-\!\varrho_{\text{min}}} {\epsilon}\right) \!N_R^{\vartheta}\right)\right), $$ where $\mathcal{I}_{\text{eff}}$ represents the effective number of iterations until the minimum SINR converges (likely much smaller than $\mathcal{I}_{\text{max}}$ as we experienced in our simulations).}

\section{Numerical Results} \label{sec:num_results}

\subsection{Simulation Scenario and Settings} \label{subsec:settings}
We consider a BS with $\Nah\!=\!\Nav\!=\!4$, i.e., $\Na\!=\!16$, antennas and an RIS with $\Nrh\!=\!\Nrv\!=\!8$, i.e., $\Nr\!=\!64$, elements serving $K\!=\!8$ UEs, unless otherwise stated. The RIS height is $h_{\RIS}\!=\!10$ m, while the BS height is $h_{\BS}\!=\!10.27$ m. The two-dimensional location of
the RIBS is $(0,\, 0)$, where the unit is meters, while the UEs are uniformly distributed at random in the 200 m $\times$ 200 m area that extends from $(0,\,-100)$ to $(200,\, 100)$. We assume that UE's height is 1.5~m, and the minimum 3D distance between the RIBS and any UE is 20 m. The carrier frequency and the bandwidth are $f_c\!=\!1.9$ GHz and $B\!=\!20$ MHz, respectively. The large-scale fading coefficients $\{\beta_{k}\}$ are modelled according to~\cite{3GPP_38901_model}. The channel correlation matrices $\{ \bR_{k} \}$ are generated by using the \textit{local scattering} model~\cite{Demir2021} assuming half-wavelength spaced RIS elements, and jointly Gaussian angular distributions of the multipath components around the nominal azimuth and elevation angles. The random variations in the azimuth and elevation angles are assumed to be independent, and the corresponding angular standard deviations (ASDs) are equal to $15^{\circ}$, which represents strong spatial channel correlation. 
We set $P^{\text{dl}}_{\text{tx,max}}\!=\!40$ W, the UL pilot power $\etap_k\!=\!\eta_k^{\text{u}}\!=200$ mW, $\forall k$. Finally, we assume a noise power spectral density of -174 dBm/Hz, with a 5 dB noise figure of the receiver.

\subsection{Performance Assessment}\label{subsec:performance}
\begin{figure}[!t]
	\centering
	\includegraphics[width=.9\columnwidth]{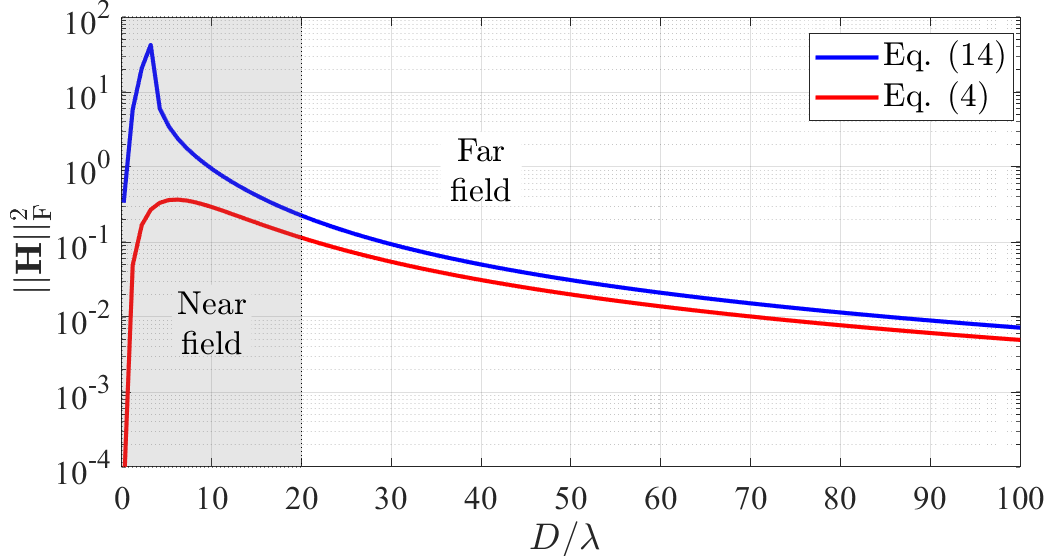}
    \vspace*{-3mm}
	\caption{{Squared Frobenius norm of the BS-to-RIS channel $\bH$ obtained via~\eqref{eq:Hnear} and~\eqref{eq:H_ij_planar} versus $D/\lambda$. Here, we assume $\Na\!=\!16$, $\Nr\!=\!64$ and $\alpha\!=\!\pi/6$. The Fraunhofer distance corresponds to $D\!=\!20\lambda.$}}
	\label{Fig:NearandFarField}
    \vspace*{-3mm}
\end{figure}

Firstly, we compare {the BS-to-RIS channel gain obtained by using the channel models described in Section \ref{subsec:channel-model}. Fig.~\ref{Fig:NearandFarField} shows the squared Frobenius norm of the channel matrix $\bH$ as a function of $D/\lambda$, i.e., the distance from the RIS to the BS normalized by the wavelength. $\bH$ given by~\eqref{eq:Hnear} captures the near-field effects and generalizes~\eqref{eq:H_ij_planar}, which is only valid for far-field communications. Indeed, we observe that~\eqref{eq:H_ij_planar} significantly underestimates the channel gain in the near-field zone, that is for values of $D/\lambda\!<\!20$, more than one order of magnitude as compared to the channel gain accurately computed via~\eqref{eq:Hnear}. In the far-field zone, i.e., $D/\lambda\!>\!20$, the gap between the two channel gains reduces, eventually differing by a multiplicative constant. Importantly, in the considered simulation scenario, we observe that $\norm{\bH}^2_{\text{F}}$ is maximum in correspondence of $D \!=\! 3.21 \lambda$ which gives the optimal relative distance between RIS and BS. Hence, in order to achieve the most positive benefits on the SE, in all the subsequent simulations, we consider $D\!=\!3.21\lambda$.\footnote{\color{blue}The SE does depend, among others, on the amount of signal energy the RIS is able to reflect, which is proportional to the energy the RIS is able to capture from the BS signal transmission. In turn, the energy of the signal received at the RIS is proportional to the BS transmit power and the BS-to-RIS channel gain, i.e, $\norm{\bH}^2_{\text{F}}$.}}%

\begin{figure}[!t]
	\centering
	\includegraphics[width=.9\columnwidth]{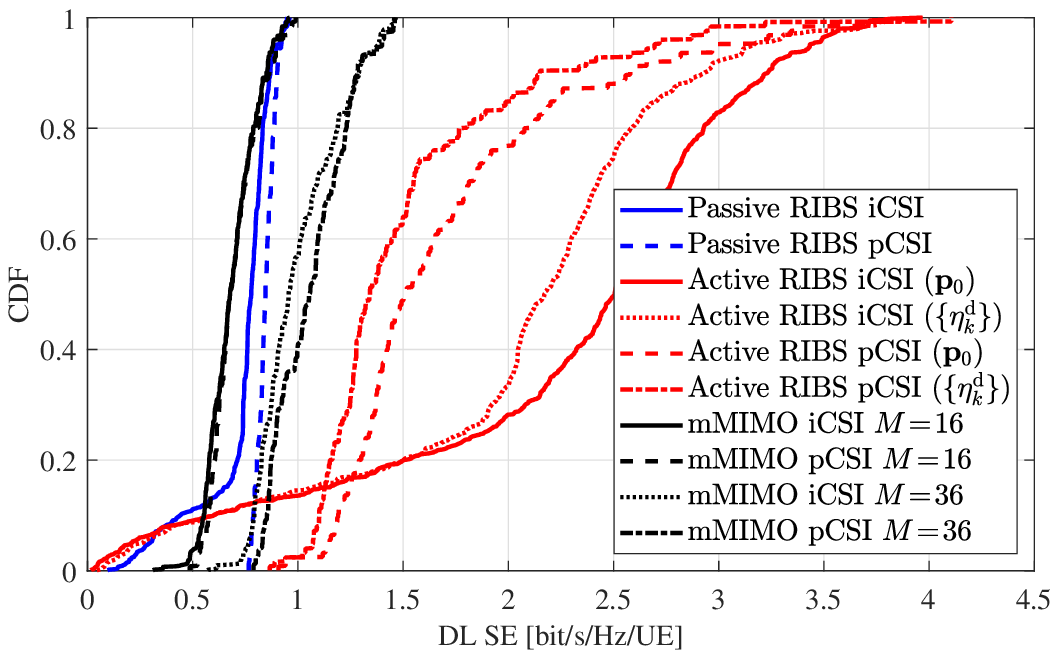}
    \vspace*{-3mm}
	\caption{CDF of the DL SE per UE achieved with MR, under pCSI and iCSI assumption. Performance comparison include RIBS with passive and active RIS and co-located MIMO (mMIMO) with $M$ antennas. Here, $K\!=\!8$, $\Na\!=\!16$, $\Nr\!=\!64$, $\alpha\!=\!\pi/6$ and $\amax\!=\!5$.}
	\label{Fig:MRT_pCSI_ipCSI}
    \vspace*{-3mm}
\end{figure}

\begin{figure}[!t]
	\centering
	\includegraphics[width=.9\columnwidth]{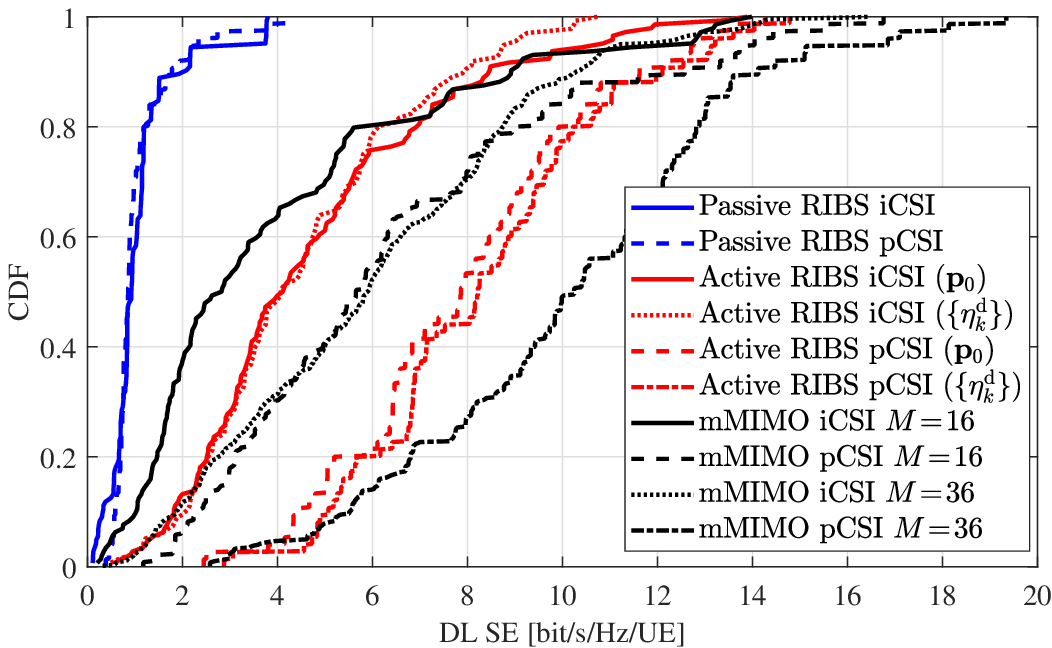}
    \vspace*{-3mm}
	\caption{CDF of the DL SE per UE achieved with MMSE, under pCSI and iCSI assumption. Performance comparison include RIBS with passive and active RIS and co-located MIMO (mMIMO) with $M$ antennas. Here, $K\!=\!8$, $\Na\!=\!16$, $\Nr\!=\!64$, $\alpha\!=\!\pi/6$ and $\amax\!=\!5$.}
	\label{Fig:MmSE_pCSI_ipCSI}
    \vspace*{-3mm}
\end{figure}

In Figs.~\ref{Fig:MRT_pCSI_ipCSI} and~\ref{Fig:MmSE_pCSI_ipCSI}, we report the cumulative distribution function
(CDF) of the DL SE per UE achieved by the proposed RIBS equipped with either a passive or an active RIS, and using either MR or MMSE precoding, respectively. Moreover, we consider both the SE achieved under perfect CSI (pCSI) and imperfect CSI (iCSI) assumption, obtained from~\eqref{eq:DL_SE_PCSI} and~\eqref{eq:DL_SE_GAIDED}, respectively. 
As for the active RIBS optimization discussed in Section \ref{sec:optimization}, we show the results obtained by using both the initialization strategies for Algorithm~\eqref{alg:alternate-optimization} proposed in~\eqref{eq:p-adjusting:active} and \eqref{eq:eta-adjusting:active}, {for which we used the notation ($\bp_0$) and ($\{\eta_k^d\}$) in the figure legend, respectively.} The former determines a set of
feasible RIS phase shifts and power control coefficients, for a given initial value of $\varepsilon$, by fine-tuning the initial phase shift vector $\bp_0$ so as to meet the power constraint at the RIS. The latter determines a set of feasible RIS phase shifts and power control coefficients, for a given initial value of $\varepsilon$, by fine-tuning the initial vector of the power control coefficients $\bEtad$ so as to meet the power constraint at the RIS.
The SE achieved by the RIBS is compared with the SE achievable by a co-located mMIMO BS with $M$ antennas, available overall DL power budget being equal. For the performance comparison to be fair, we consider that mMIMO implements the max-min fairness power control as described in~\cite[Section 7.1.1]{massivemimobook}. {Moreover, notice that the uplink training duration is $\tp\!=\!K$ for the mMIMO case and $\tp\!=\!K(\Nr+1)$ for the RIBS case.}

Fig. \ref{Fig:MRT_pCSI_ipCSI} shows that, in case of pCSI, the active RIBS outperforms the mMIMO system with 16 and 36 antennas, while the passive RIBS outperforms the mMIMO with 16 antennas. This significant SE improvement is justified by a higher number of degrees of freedom provided by the active RIBS. Under the assumption of iCSI, unfortunately, the minimum SE maximization strategy loses its effectiveness if MR is employed at the RIBS.
Let us recall that, in case of iCSI, the SE expression involved in the proposed joint optimization problem is given by~\eqref{eq:DL_SE_ICSI}, computed upon the information available at the RIBS. While, the SE values reported in all the figures are obtained by inserting the optimal solutions of Algorithm~\ref{alg:alternate-optimization} into the SE expression in~\eqref{eq:DL_SE_GAIDED}. The SE expressions~\eqref{eq:DL_SE_ICSI} and~\eqref{eq:DL_SE_GAIDED} clearly coincide under the assumption of pCSI. Hence, the negative impact of the channel estimation error on the SE is amplified when using MR, and especially in presence of the dynamic noise introduced by the active RIBS. The ineffectiveness of the max-min fairness SE optimization in case of iCSI and MR motivates the long tail of the SE CDF in Fig.~\ref{Fig:MRT_pCSI_ipCSI}, and the inability of equalizing the SE achieved by the UEs with better conditions in favor of the UEs with worse channel conditions.

On the other hand, MMSE precoding is able to overcome this limitation thanks to its capability of mitigating the interference, including that caused by the channel estimation error and by the dynamic noise in case of active RIBS. This aspect is clearly observable in Fig.~\ref{Fig:MmSE_pCSI_ipCSI}, which shows the DL achievable SE assuming MMSE precoding.
Notably, Fig.~\ref{Fig:MmSE_pCSI_ipCSI} also reveals that both mMIMO and active RIBS can largely benefit from a more sophisticated precoding strategy, with the active RIBS uniformly outperforming mMIMO with 16 antennas and providing the same 95\%-likely SE achieved by mMIMO with 36 antennas, namely with more than the double of the active antenna elements of the RIBS. %
{Hence, performance being equal, the proposed RIBS architecture, equipped with an active RIS, provides a significant saving in terms of power consumption with respect to co-located mMIMO, since the required extra power for an active RIS reflective element is at most 10 mW~\cite{Liu2021}, while the required power for RF chains and amplifiers of an active array is about 1 W per antenna element~\cite[Section 5.4]{massivemimobook}. This leads, in the considered scenario, to the RIBS achieving a power saving of about 20 W compared to mMIMO, basically determined by the difference in the number of antenna elements at the respective active arrays, that is $M \!-\! \Na$.} %
Despite the use of MMSE precoding, the passive RIBS 
is not able to improve its performance. The latter demonstrates how crucial the fine-tuning of the RIS element amplification factor is to enhance the DL SE. As a final comment, we observe that the initialization strategies for Algorithm~\eqref{alg:alternate-optimization} proposed in~\eqref{eq:p-adjusting:active} and \eqref{eq:eta-adjusting:active} perform equally well for MMSE, while the strategy in~\eqref{eq:eta-adjusting:active} pays off for MR. This result further demonstrates that, even at the first iteration of Algorithm~\eqref{alg:alternate-optimization} is more beneficial to constrain the transmit powers than the RIS phase-shifts, which are the only countermeasures to the interference, instead of the opposite. Conversely, the SE achieved by MMSE is less sensitive to the initial feasible set of the optimization problem.  

\begin{figure}[!t]
	\centering
	\includegraphics[width=.9\columnwidth]{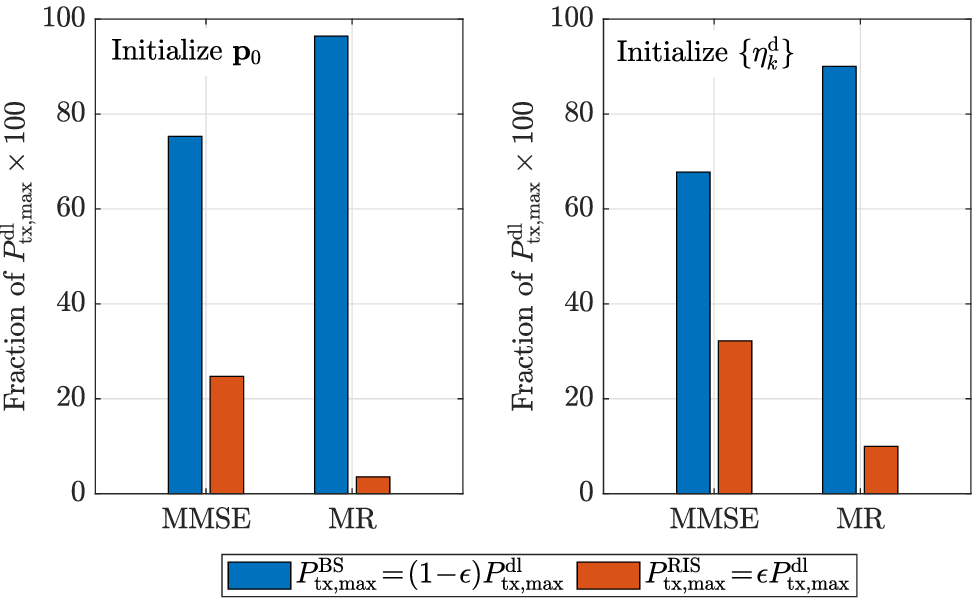}
    \vspace*{-3mm}
	\caption{Power split between RIS and BS in an active RIBS, assuming MMSE and MR with the two initialization strategies proposed in~\eqref{eq:p-adjusting:active} and \eqref{eq:eta-adjusting:active}. The simulation settings are the same as those used to obtain Figs.~\ref{Fig:MRT_pCSI_ipCSI} and \ref{Fig:MmSE_pCSI_ipCSI}.}
	\label{Fig:Transmit_power_budget}
    \vspace*{-3mm}
\end{figure}

In Fig. \ref{Fig:Transmit_power_budget}, we show how the transmit power is split between BS and RIS in an active RIBS, both for MMSE and MR, and considering the two aforementioned initialization strategies. The simulation settings are the same as those used to produce the results in Figs.~\ref{Fig:MRT_pCSI_ipCSI} and \ref{Fig:MmSE_pCSI_ipCSI}. Regardless of the precoding scheme, a larger share of power is allocated at the BS rather than at the RIS, as expected. Interestingly, more power is allocated at the RIS in MMSE compared to that in the case of MR. Indeed, MMSE precoding aims at minimizing the interference, which can be achieved only by fully leveraging the degrees of freedom provided by the optimization of the RIS phase-shifts. Conversely, MR aims at maximizing the power of the desired signal which can be attained by employing as much ``active'' transmit power as possible.

\begin{figure}[!t]
	\centering
	\includegraphics[width=.9\columnwidth]{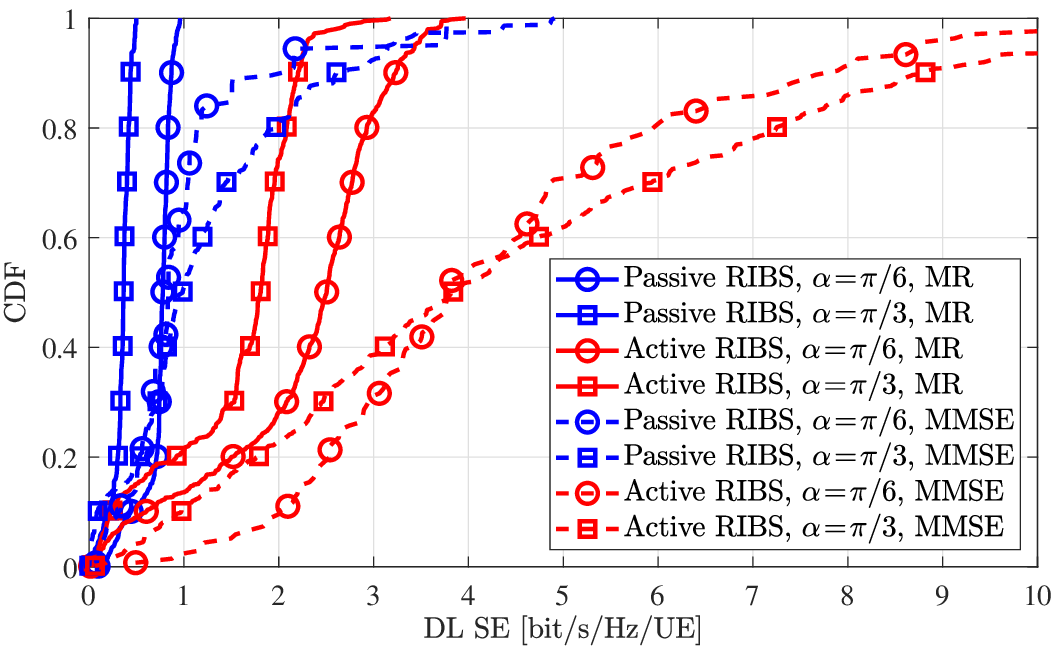}
    \vspace*{-3mm}
	\caption{CDF of the DL SE per UE achieved with MMSE and MR, under iCSI assumption, for different values of downtilt angle of the BS's UPA with respect to the RIS's UPA. Here, $K\!=\!8$, $\Na\!=\!16$, $\Nr\!=\!64$ and $\amax\!=\!5$. }
	\label{Fig:MmSE_MRT_ipCSI_alpha}
    \vspace*{-3mm}
\end{figure}

In Fig.~\ref{Fig:MmSE_MRT_ipCSI_alpha}, we investigate the impact of the downtilt angle of the BS's UPA with respect to the RIS's UPA plane, i.e., $\alpha$, on the performance achieved by the proposed RIBS. The achievable DL SE shown in Fig.~\ref{Fig:MmSE_MRT_ipCSI_alpha} for MR and MMSE is the highest obtained between the two initialization strategies for Algorithm~\eqref{alg:alternate-optimization} proposed in~\eqref{eq:p-adjusting:active} and \eqref{eq:eta-adjusting:active}. 
We observe that the SE obtained with $\alpha\!=\!\pi/6$ is larger than that obtained with $\alpha=\pi/3$. This result aligns with our expectations as at $\pi/6$ the transmit UPA surface of the BS that intercepts the UPA of the RIS is larger, hence more transmit energy can be captured by the RIS, and, in turn, reflected towards the UEs. As we increase the value of $\alpha$ approaching the BS's UPA downtilt angle value of $\pi/2$,  the SE degrades due to the smaller energy of the near-field signal impinging on the RIS.%

\begin{figure}[!t]
	\centering
	\includegraphics[width=.8\columnwidth]{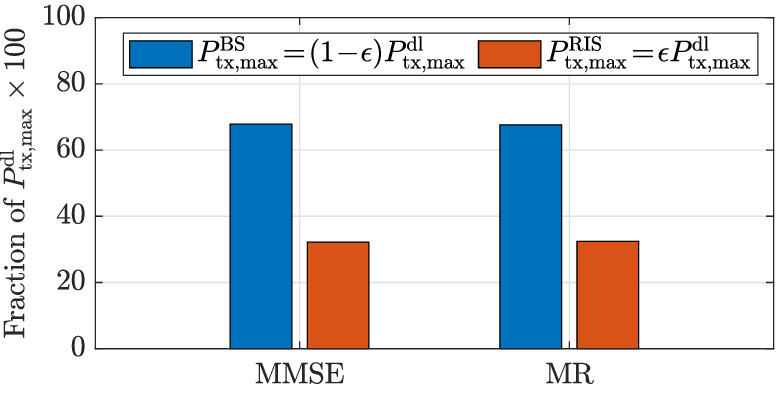}
    \vspace*{-3mm}
	\caption{Power split between RIS and BS in an active RIBS, assuming MMSE and MR, with $\alpha\!=\!\pi/3$. 
    Settings are the same as those used to obtain Fig.~\ref{Fig:MmSE_MRT_ipCSI_alpha}.}
	\label{Fig:Transmit_power_budge_pi3}
    \vspace*{-3mm}
\end{figure}

In Fig.~\ref{Fig:Transmit_power_budge_pi3}, we show how the transmit power is split between BS and RIS in an active RIBS, both for MMSE and MR, considering $\alpha=\pi/3$. As compared to the results shown in Fig.~\ref{Fig:Transmit_power_budget} for $\alpha=\pi/6$, we notice a more uniform distribution of the transmit power between RIS and BS, especially for MR. Since the energy of the signal impinging on the RIS is limited by an inappropriate choice of the BS's downtilt angle (i.e., the transmit power radiated by the BS is lost in space), then it is reasonable not to waste too much DL power at the BS and instead invest a larger share of it at the RIS.%

\begin{figure}[!t]
	\centering
	\includegraphics[width=.82\columnwidth]{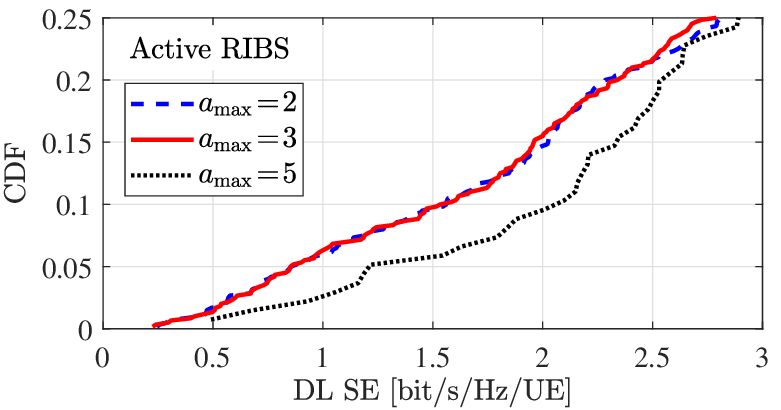}
    \vspace*{-3mm}
	\caption{CDF of the DL SE per UE achieved by the active RIBS, under iCSI assumption, assuming MMSE and different values of the RIS element amplification factor, i.e., $\amax$. Here, $\Na\!=\!16$, $\Nr\!=\!64$, $K\!=\!8$ and $\alpha\!=\!\pi/6$.}
	\label{Fig:MmSE_ipCSI_amax}
    \vspace*{-3mm}
\end{figure}

Now, we investigate the impact of the maximum value of the RIS element amplification factor, i.e., $\amax$, on the performance achieved by the active RIBS. In Fig.~\ref{Fig:MmSE_ipCSI_amax}, we show the CDF of the DL SE per UE as $\amax$ varies. We observe a negligible performance variation for $\amax\!=\!2$ and $\amax\!=\!3$. Conversely, with $\amax\!=\!5$ we obtain an increase of about 32\% and 42\% in terms of 90\%-likely and 95\%-likely SE, respectively. Increasing the RIS element amplification factor provides more degrees of freedom in the RIS phase-shift optimization, thereby resulting on higher SE values at the lower percentiles of the CDF. (Let us recall that the proposed optimization strategy aims at maximizing the minimum per-UE SE in the system.) On the other hand, increasing $\amax$ requires allocating more power at the RIS than at the BS, which, for sufficiently large values of $\amax$, may eventually lead to a SE degradation.
Indeed, the optimal power split value resulting from Algorithm~\ref{alg:alternate-optimization} consistently suggests that allocating more ``active'' transmit power at the BS than ``reflective'' power at the RIS is more beneficial, as shown in Figs.~\ref{Fig:MmSE_MRT_ipCSI_alpha} and~\ref{Fig:Transmit_power_budge_pi3}. 

\begin{figure}[!t]
	\centering
	\includegraphics[width=.98\columnwidth]{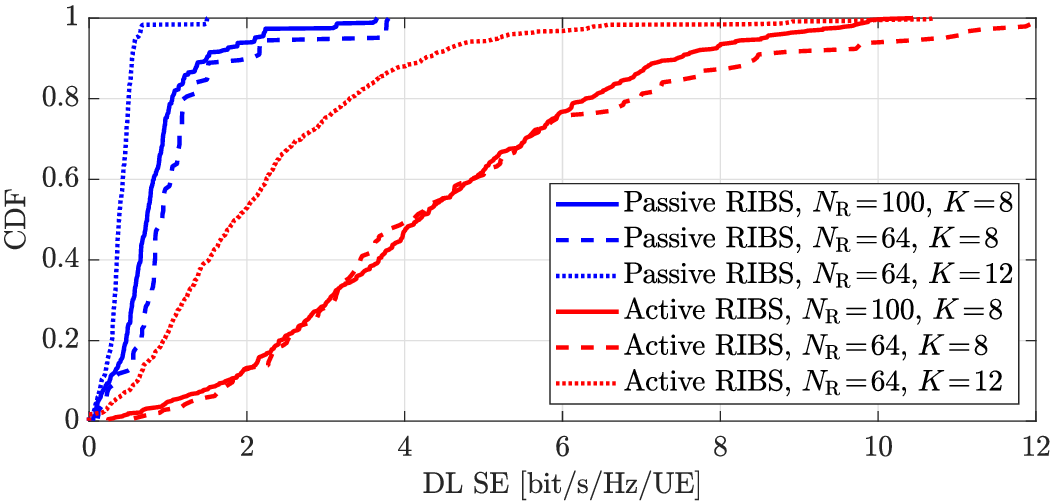}
    \vspace*{-3mm}
	\caption{CDF of the DL SE per UE achieved by the RIBS, under iCSI assumption, assuming MMSE and different values for $K$ and $\Nr$. Here, $\Na\!=\!16$, $\alpha\!=\!\pi/6$ and $\amax\!=\!5$.}
	\label{Fig:MmSE_ipCSI_NR_K}
    \vspace*{-3mm}
\end{figure}

In Fig.~\ref{Fig:MmSE_ipCSI_NR_K}, we investigate the impact of the number of RIS elements, $\Nr$, and of the number of UEs, $K$, on the achievable DL SE per UE. As we may observe, the increase of $\Nr$, from 64 to 100, degrades the SE achieved by the passive RIBS. This degradation is due to {two aspects: $(i)$ the channel estimation accuracy which depends on the $\Na$ active antennas and is negatively influenced by the increase of $\Nr$ and $(ii)$} additional interference caused by an increased number of RIS reflections that becomes harder to mitigate despite the use of MMSE precoding. On the other hand, the increase of $\Nr$, from 64 to 100, has a negligible impact on SE achieved by the active RIBS which can better handle the additional interference via RIS phase-shift optimization by resorting to degrees of freedom provided by the non-unitary RIS element amplification factor. 
As expected, the increase in the number of UEs, from 8 to 12, uniformly lowers the SE per UE, due to the higher amount of multi-user interference in the network. This interference may possibly be suppressed by increasing the number of antennas at the BS, $\Na$, to make MMSE precoding more effective.

\section{{Conclusions and Future Works}}%
\label{sec:conclusions}

We proposed a reconfigurable intelligent base station (RIBS) consisting of a small planar antenna array illuminating a built-in RIS in its near field, along with a multiuser beamforming design achieving comparable performance with that of fully-digital co-located mMIMO systems with larger active arrays, hence {resulting in a reduced hardware complexity and thereby leading to a more cost-effective and energy-efficient solution.} The key idea is to serve the UEs via reflection by the RIS upon joint optimization of the RIS phase-shifts and DL transmit powers to achieve high gain and high directivity of the beams, and to maximize the minimum SE in the network. 

We provided a detailed analysis including: $(i)$ the derivation of a closed-form expression for the BS-to-RIS channel that captures the near-field effects; $(ii)$ the derivation of an achievable DL SE expression accounting for both active and passive RIS and imperfect CSI at the RIBS; $(iii)$ an alternate optimization to jointly determine the RIS phase-shifts, the DL powers, and the fraction of power split between BS and RIS that maximizes the minimum SE in the network; and $(iv)$ insights on favorable propagation and channel hardening concerning the link from the BS to the UE via the RIS. 

{The key findings of this study are summarized as follows:
\begin{itemize}
    \item[-] the RIBS with an active RIS greatly outperforms conventional massive MIMO systems, number of RF chains being equal, showing up to a 50\% improvement in SE. 
    \item[-] User fairness is especially guaranteed when adopting MMSE precoding thanks to its capability of mitigating the interference, including that caused by the channel estimation error and by the active RIS dynamic noise. The fairness optimization resulted in a 30\% gain in SE uniformity across UEs, ensuring that no single user experiences a significant performance drop.
    \item[-] RIBS with active RIS was shown to achieve better SE due to its ability to both amplify and phase-shift the incoming signals, whereas the passive RIS, while still effective, showed more modest improvements.
    \item[-] Our analysis demonstrated that channel hardening and favorable propagation, key advantages of massive MIMO systems, can also be approached with the RIBS setup. However, they do not benefit from the use of an RIS.
    \item[-] A proper and accurate characterization of the near-field channel between RIS and BS enables an optimal designing of the RIBS by selecting the best relative distance between BS and RIS that yields the highest channel gain.  
\end{itemize}
The proposed RIBS structure can be reasonably appealing for beyond-5G networks, where high performance is required with reduced energy consumption and lower hardware costs.
The RIBS system is particularly suited for dense urban environments or areas with heavy signal blockage, offering improved control over the wireless environment through RIS technology.}

{Future research could extend the RIBS concept to explore applications such as user localization and tracking, leveraging RIS for precise positioning, thereby enhancing the accuracy of location-based services. Additionally, further exploration of multiuser scenarios, as well as RIS deployment in different environmental settings, could provide deeper insights into the practical deployment and optimization of RIS-aided communication systems. For instance, the proposed RIBS architecture would only be able to serve UEs within a 180-degree range. A possible setup enabling to serve UEs with a 360-degree range would consist in deploying two RIBSs back to back or, as an alternative, a \textit{simultaneously transmitting and reflecting reconfigurable intelligent surface} (STAR-RIS)~\cite{Liu2021}, which is able to reflect and refract (transmit), thus serving UEs on both sides of the transmitter. The analysis of such systems are undoubtedly an interesting research direction for future works. Finally, a valuable extension of our work would consist in investigating well-established optimization methods, such as majorization-minimization, manifold and semidefinite relaxation, for RIBS equipped with active RIS, as their performance and effectiveness in this context remain an open area of research~\cite{Peng2024}}.

\appendix[BS-to-RIS Near-Field Channel]
\label{subsec:appendix-nearfield}
We provide the analytical steps to derive the closed-form expression~\eqref{eq:channel-gain:near-field}. Here, we assume that the received signal's phase variations are negligible over the RIS area.
Let us begin with considering the expression of the complex-valued channel in the $xy$-plane from an arbitrary antenna of the BS's UPA located in $(x_\tA, y_\tA, z_\tA)$ to an arbitrary element of the RIS located in $(x_\tR, y_\tR, z_\tR)$, and placed in its \textit{radiative} near field~\cite[Appendix A]{Bjornson2020NearField}. For the sake of brevity, we herein omit the indices $m$ and $n$ that identify the BS antenna and the RIS element, respectively. 
We consider the following integral:
\begin{equation}
\label{eq:I}
\!\!\!\!I \!=\!\! \int_{x_\tR-\Drh/2}^{x_\tR+\Drh/2}\!\!\int_{y_\tR-\Drv/2}^{y_\tR+\Drv/2}\!\!\int_{B_\tx}^{A_\tx}\!\!\int_{B_\ty}^{C_\ty} \!\!\!f(x,y,\bar{x},\bar{y})\,d\bar{x}\,d\bar{y}\,dx\,dy \,,
\end{equation}
where 
\begin{equation}
f(x,y,\bar{x},\bar{y}) = \frac{z_\tA\left[(y-\bar{y})^2+z_\tA^2\right]}{\left[(x-\bar{x})^2+(y-\bar{y})^2+z_\tA^2\right]^{5/2}}\, ,
\end{equation}
denotes the channel gain along the $z$-axis, and
\begin{align*}
A_\tx \!&=\! \frac{\Delta_{\text{c}}\sin(\gamma\!+\!\Delta_\Phi)}{2\sin(\Delta_{\Phi}/2)} \, , & & B_\tx \!=\!\frac{\Delta_{\text{c}}\sin\gamma}{2\sin(\Delta_{\Phi}/2)}\, , \\ 
B_\ty \!&=\! y_\tR \!-\! y_\tA \!-\! (\Da/2) \cos\alpha \, , & &
C_\ty \!=\! y_\tR\!-\!y_\tA \!+\! (\Da/2) \cos\alpha \, ,
\end{align*}
with $\gamma \!=\! (1/D)\arctan\left(x_\tA \!-\!\Da/2\!-\!x_\tR\right)$, $\Delta_{\text{c}}$ being the rope of the extremes $A_\tx$ and $B_\tx$, and $\Delta_\Phi$ being the angle subtended by this rope.
Upon changing the integration variables $\hat{x} \!=\!x-\bar{x}$ and $\hat{y}\!=\!y-\bar{y}$, similar to~\cite[Eq.\,(4)]{DiMurro2020}, \eqref{eq:I} can be written as
\begin{equation}
\label{eq:I2}
\!\! I \!=\!\!\! \int_{x_\tR-\Drh/2}^{x_\tR+\Drh/2}\!\!\!\int_{y_\tR-\Drv/2}^{y_\tR+\Drv/2}\!\!\!\int_{x-A_\tx}^{x-B_\tx}\!\!\!\int_{y-C_\ty}^{y-B_\ty} \!\!\! f(\hat{x},\hat{y})\,d\hat{x}\,d\hat{y}\,dx\,dy \, .
\end{equation}
By using the same approach as in~\cite{DiMurro2020}, we rotate the integration domain by 90 degrees. This step allows us to reformulate~\eqref{eq:I2} as the sum of nine two-dimensional integrals, as shown in~\eqref{eq:I4} at the top of the this page.
\begin{figure*}\small
    \begin{align}
    \label{eq:I4}
	I &=\! \int_{x_{1}}^{x_{2}}\int_{y_{1}}^{y_{2}}(\hat{x}-x_{1})~(\hat{y}-y_{1})~f(\hat{x},\hat{y})~d\hat{x}~d\hat{y} \!+\! \int_{x_{1}}^{x_{2}}\int_{y_{2}}^{y_{3}}(\hat{x}-x_{1})~\Delta^\ty	\, f(\hat{x},\hat{y})~d\hat{x}~d\hat{y} \!+\! \int_{x_{2}}^{x_{3}}\int_{y_{1}}^{y_{2}} \Delta^\tx~(\hat{y}-y_{1})~f(\hat{x},\hat{y})~d\hat{x}~d\hat{y} \nonumber \\    &\quad\;+\!\int_{x_{2}}^{x_{3}}\int_{y_{2}}^{y_{3}}\Delta^\tx~\Delta^\ty~f(\hat{x},\hat{y})~d\hat{x}~d\hat{y} \!+\! \int_{x_{1}}^{x_{2}}\int_{y_{3}}^{y_{4}}(\hat{x}-x_{1})~(y_{4}-\hat{y})	f(\hat{x},\hat{y})~d\hat{x}~d\hat{y} \!+\! \int_{x_{2}}^{x_{3}}\int_{y_{3}}^{y_{4}}\Delta^\tx~(y_{4}-\hat{y})~f(\hat{x},\hat{y})~d\hat{x}~d\hat{y} \nonumber \\
	&\quad\;+\!\int_{x_{3}}^{x_{4}}\int_{y_{1}}^{y_{2}}(x_{4}-\hat{x})\,(\hat{y}-y_{1})\,f(\hat{x},\hat{y}) \,d\hat{x}\,d\hat{y} \!+\! \int_{x_{3}}^{x_{4}}\int_{y_{2}}^{y_{3}}(x_{4}-\hat{x}) \,\Delta^\ty\,f(\hat{x},\hat{y})\,d\hat{x}\,d\hat{y} \!+\! \int_{x_{3}}^{x_{4}}\int_{y_{3}}^{y_{4}}(x_{4}-\hat{x})\,(y_{4}-\hat{y})\,f(\hat{x},\hat{y})\,d\hat{x}\,d\hat{y}
\end{align}
\hrulefill
\vspace*{-3mm}
\end{figure*}
Notice that $x_1,\ldots,x_4$ and $y_1,\ldots,y_4$ in~\eqref{eq:I4} were defined in~\Secref{subsec:channel-model} for an arbitrary BS antenna $m$ and RIS element $n$.
These integrals can be computed in closed form yielding~\eqref{eq:channel-gain:near-field}. 

\bibliographystyle{IEEEtran}
\bibliography{IEEEabrv,biblio}

\end{document}

%% file: my-macro.tex
\usepackage{mathtools} 
\usepackage{xspace}

\newcommand{\eqrefs}[2]{(\ref{#1})-(\ref{#2})}
\newcommand{\Figref}[1]{Fig.~\ref{#1}}

\newcommand{\Secref}[1]{Section~\ref{#1}}

\newcommand{\ba}{\mathbf{a}}

\newcommand{\bh}{\mathbf{h}}

\newcommand{\bp}{\mathbf{p}}

\newcommand{\br}{\mathbf{r}}

\newcommand{\bv}{\mathbf{v}}
\newcommand{\bw}{\mathbf{w}}
\newcommand{\by}{\mathbf{y}}

\newcommand{\bz}{\mathbf{z}}

\newcommand{\bA}{\mathbf{A}}
\newcommand{\bB}{\mathbf{B}}

\newcommand{\bG}{\mathbf{G}}
\newcommand{\bH}{\mathbf{H}}
\newcommand{\bI}{\mathbf{I}}

\newcommand{\bP}{\mathbf{P}}

\newcommand{\bR}{\mathbf{R}}

\newcommand{\bW}{\mathbf{W}}

\newcommand{\bvphi}{\boldsymbol{\varphi}}

\newcommand{\bEtad}{\boldsymbol{\eta}^{\text{d}}}

\newcommand{\bzero}{\boldsymbol{0}}

\newtheorem{lemma}{Lemma}

\newcommand{\real}[1]{\Re\left\{#1\right\}}

\newcommand{\diag}{\text{diag}}

\newcommand{\herm}{^{\mathsf{H}}}
\newcommand{\trans}{^\mathsf{T}}
\newcommand{\inv}{^{-1}}

\DeclareMathOperator{\tr}{tr}
\DeclareMathOperator{\E}{\mathsf{E}}

\newcommand{\EX}[1]{\E\left\{{#1}\right\}}

\newcommand{\euler}{\mathrm{e}}

\newcommand*{\imagj}{\mathsf{j}} 

\newcommand{\norm}[1]{{ \left\Vert #1 \right\Vert }}

\newcommand{\mC}{\mathbb{C}}

\newcommand{\tc}{\tau_{\mathrm{c}}}
\newcommand{\tp}{\tau_{\mathrm{p}}}
\newcommand{\td}{\tau_{\mathrm{d}}}
\newcommand{\tu}{\tau_{\mathrm{u}}}

\newcommand{\CG}[2]{\mathcal{CN}\left({#1},{#2}\right)}

\newcommand{\bYp}{\mathbf{Y}^{\text{p}}}
\newcommand{\byp}{\mathbf{y}^{\text{p}}}
\newcommand{\bzp}{\mathbf{z}^{\text{p}}}
\newcommand{\bNp}{\mathbf{N}^{\text{p}}}
\newcommand{\bnp}{\mathbf{n}^{\text{p}}}
\newcommand{\wtbNp}{\wt{\mathbf{N}}^{\text{p}}}
\newcommand{\wtbnp}{\wt{\mathbf{n}}^{\text{p}}}
\newcommand{\etap}{\eta^{\text{p}}}
\newcommand{\etad}{\eta^{\text{d}}}
\newcommand{\up}{^{\text{u}}}

\newcommand{\pCSI}{^{\text{pCSI}}}
\newcommand{\iCSI}{^{\text{iCSI}}}
\newcommand{\Na}{N_{\text{A}}}
\newcommand{\Nr}{N_{\text{R}}}
\newcommand{\BS}{\text{BS}}
\newcommand{\UE}{\text{UE}}
\newcommand{\RIS}{\text{RIS}}
\newcommand{\Da}{\Delta_{\text{A}}}
\newcommand{\Dr}{\Delta_{\text{R}}}
\newcommand{\Dav}{\Da^{\text{v}}}
\newcommand{\Dah}{\Da^{\text{h}}}
\newcommand{\Drv}{\Dr^{\text{v}}}
\newcommand{\Drh}{\Dr^{\text{h}}}
\newcommand{\Nav}{N_{\text{A}}^{\text{v}}}
\newcommand{\Nah}{N_{\text{A}}^{\text{h}}}
\newcommand{\Nrv}{N_{\text{R}}^{\text{v}}}
\newcommand{\Nrh}{N_{\text{R}}^{\text{h}}}
\newcommand{\dav}{d_{\text{A}}^{\text{v}}}
\newcommand{\dah}{d_{\text{A}}^{\text{h}}}
\newcommand{\drv}{d_{\text{R}}^{\text{v}}}
\newcommand{\drh}{d_{\text{R}}^{\text{h}}}
\newcommand{\da}{d_{\text{A}}}
\newcommand{\dr}{d_{\text{R}}}
\newcommand{\tR}{{\text{R}}}
\newcommand{\tA}{{\text{A}}}
\newcommand{\tx}{{\text{x}}}
\newcommand{\ty}{{\text{y}}}
\newcommand{\tC}{{\text{c}}}

\newcommand{\setX}{\mathcal{X}}
\newcommand{\setY}{\mathcal{Y}}
\newcommand{\amax}{a_{\text{max}}}

%% file: fig/RIBS.tex
\tikzset {_q7xzgkamo/.code = {\pgfsetadditionalshadetransform{ \pgftransformshift{\pgfpoint{0 bp } { 0 bp }  }  \pgftransformrotate{-23 }  \pgftransformscale{2 }  }}}
\pgfdeclarehorizontalshading{_3jcu0cz6i}{150bp}{rgb(0bp)=(1,0,0);
rgb(37.5bp)=(1,0,0);
rgb(43.75bp)=(1,1,0);
rgb(50bp)=(0.02,0.76,1);
rgb(56.25bp)=(1,1,0);
rgb(62.5bp)=(1,0,0);
rgb(100bp)=(1,0,0)}

  
\tikzset {_6twq0hqa4/.code = {\pgfsetadditionalshadetransform{ \pgftransformshift{\pgfpoint{0 bp } { 0 bp }  }  \pgftransformrotate{0 }  \pgftransformscale{2 }  }}}
\pgfdeclarehorizontalshading{_h1f5ktwi3}{150bp}{rgb(0bp)=(1,0.99,0.92);
rgb(37.5bp)=(1,0.99,0.92);
rgb(62.5bp)=(0.95,0.85,0.21);
rgb(100bp)=(0.95,0.85,0.21)}

  
\tikzset {_gl1m40bbq/.code = {\pgfsetadditionalshadetransform{ \pgftransformshift{\pgfpoint{0 bp } { 0 bp }  }  \pgftransformrotate{0 }  \pgftransformscale{2 }  }}}
\pgfdeclarehorizontalshading{_4i90ziygl}{150bp}{rgb(0bp)=(1,1,1);
rgb(37.5bp)=(1,1,1);
rgb(62.5bp)=(0.82,0.01,0.11);
rgb(100bp)=(0.82,0.01,0.11)}

  
\tikzset {_bm6za6y7b/.code = {\pgfsetadditionalshadetransform{ \pgftransformshift{\pgfpoint{0 bp } { 0 bp }  }  \pgftransformrotate{0 }  \pgftransformscale{2 }  }}}
\pgfdeclarehorizontalshading{_96ealr2rs}{150bp}{rgb(0bp)=(1,1,1);
rgb(37.5bp)=(1,1,1);
rgb(62.5bp)=(0.39,0.58,0.76);
rgb(100bp)=(0.39,0.58,0.76)}
\tikzset{every picture/.style={line width=0.75pt}} 

\begin{tikzpicture}[x=0.75pt,y=0.75pt,yscale=-1,xscale=1]

\draw    (271.59,273.83) -- (182.86,308.94) ;
\draw [shift={(181,309.67)}, rotate = 338.41] [color={rgb, 255:red, 0; green, 0; blue, 0 }  ][line width=0.75]    (10.93,-3.29) .. controls (6.95,-1.4) and (3.31,-0.3) .. (0,0) .. controls (3.31,0.3) and (6.95,1.4) .. (10.93,3.29)   ;
\draw    (283.79,273.83) -- (461.04,310.27) ;
\draw [shift={(463,310.67)}, rotate = 191.62] [color={rgb, 255:red, 0; green, 0; blue, 0 }  ][line width=0.75]    (10.93,-3.29) .. controls (6.95,-1.4) and (3.31,-0.3) .. (0,0) .. controls (3.31,0.3) and (6.95,1.4) .. (10.93,3.29)   ;
\draw    (274.76,273.28) -- (274.76,33.91) ;
\draw [shift={(274.76,31.91)}, rotate = 90] [color={rgb, 255:red, 0; green, 0; blue, 0 }  ][line width=0.75]    (10.93,-3.29) .. controls (6.95,-1.4) and (3.31,-0.3) .. (0,0) .. controls (3.31,0.3) and (6.95,1.4) .. (10.93,3.29)   ;
\draw  [fill={rgb, 255:red, 128; green, 128; blue, 128 }  ,fill opacity=1 ] (282.68,79.98) -- (282.68,273.83) .. controls (282.68,274.75) and (280.2,275.5) .. (277.13,275.5) .. controls (274.07,275.5) and (271.59,274.75) .. (271.59,273.83) -- (271.59,79.98) .. controls (271.59,79.06) and (274.07,78.31) .. (277.13,78.31) .. controls (280.2,78.31) and (282.68,79.06) .. (282.68,79.98) .. controls (282.68,80.89) and (280.2,81.64) .. (277.13,81.64) .. controls (274.07,81.64) and (271.59,80.89) .. (271.59,79.98) ;
\draw  [fill={rgb, 255:red, 128; green, 128; blue, 128 }  ,fill opacity=1 ] (335.32,87.83) -- (330.43,84.02) -- (227.85,115.22) -- (228.34,186.86) -- (233.22,190.67) -- (335.8,159.47) -- cycle ; \draw   (227.85,115.22) -- (232.73,119.03) -- (335.32,87.83) ; \draw   (232.73,119.03) -- (233.22,190.67) ;
\path  [shading=_3jcu0cz6i,_q7xzgkamo] (335.57,159.54) -- (335.18,87.92) -- (233.15,119.17) -- (233.54,190.78) -- cycle ; 
 \draw   (335.57,159.54) -- (335.18,87.92) -- (233.15,119.17) -- (233.54,190.78) -- cycle ; 

\draw [color={rgb, 255:red, 155; green, 155; blue, 155 }  ,draw opacity=1 ]   (335.52,96.87) -- (232.62,127.81) ;
\draw [color={rgb, 255:red, 155; green, 155; blue, 155 }  ,draw opacity=1 ]   (334.73,106.62) -- (233.54,136.69) ;
\draw [color={rgb, 255:red, 155; green, 155; blue, 155 }  ,draw opacity=1 ]   (334.8,115.62) -- (233.61,145.7) ;
\draw [color={rgb, 255:red, 155; green, 155; blue, 155 }  ,draw opacity=1 ]   (334.72,125.72) -- (233.53,155.8) ;
\draw [color={rgb, 255:red, 155; green, 155; blue, 155 }  ,draw opacity=1 ]   (334.87,134.56) -- (233.68,164.64) ;
\draw [color={rgb, 255:red, 155; green, 155; blue, 155 }  ,draw opacity=1 ]   (334.87,144.2) -- (233.68,174.27) ;
\draw [color={rgb, 255:red, 155; green, 155; blue, 155 }  ,draw opacity=1 ]   (334.87,152.89) -- (233.68,182.97) ;
\draw [color={rgb, 255:red, 155; green, 155; blue, 155 }  ,draw opacity=1 ]   (241.76,116.61) -- (242.23,188.28) ;
\draw [color={rgb, 255:red, 155; green, 155; blue, 155 }  ,draw opacity=1 ]   (251.16,114.26) -- (251.63,185.93) ;
\draw [color={rgb, 255:red, 155; green, 155; blue, 155 }  ,draw opacity=1 ]   (260.8,111.21) -- (261.27,182.87) ;
\draw [color={rgb, 255:red, 155; green, 155; blue, 155 }  ,draw opacity=1 ]   (269.26,108.39) -- (269.73,180.05) ;
\draw [color={rgb, 255:red, 155; green, 155; blue, 155 }  ,draw opacity=1 ]   (288.76,102.04) -- (289.23,173.71) ;
\draw [color={rgb, 255:red, 155; green, 155; blue, 155 }  ,draw opacity=1 ]   (298.16,99.53) -- (298.63,171.19) ;
\draw [color={rgb, 255:red, 155; green, 155; blue, 155 }  ,draw opacity=1 ]   (307.32,96.24) -- (307.79,167.9) ;
\draw [color={rgb, 255:red, 155; green, 155; blue, 155 }  ,draw opacity=1 ]   (316.95,93.18) -- (317.42,164.85) ;
\draw [color={rgb, 255:red, 155; green, 155; blue, 155 }  ,draw opacity=1 ]   (326.35,90.56) -- (326.82,162.23) ;
\draw   (311.39,351.97) .. controls (311.39,350.93) and (312.23,350.09) .. (313.27,350.09) .. controls (314.3,350.09) and (315.15,350.93) .. (315.15,351.97) .. controls (315.15,353.01) and (314.3,353.85) .. (313.27,353.85) .. controls (312.23,353.85) and (311.39,353.01) .. (311.39,351.97) -- cycle ;
\draw   (317.12,332.24) .. controls (318.63,332.24) and (319.85,333.46) .. (319.85,334.96) -- (319.85,353.95) .. controls (319.85,355.45) and (318.63,356.67) .. (317.12,356.67) -- (308.94,356.67) .. controls (307.44,356.67) and (306.22,355.45) .. (306.22,353.95) -- (306.22,334.96) .. controls (306.22,333.46) and (307.44,332.24) .. (308.94,332.24) -- cycle ;

\draw   (360.26,328.61) .. controls (360.26,327.57) and (361.1,326.73) .. (362.14,326.73) .. controls (363.18,326.73) and (364.02,327.57) .. (364.02,328.61) .. controls (364.02,329.65) and (363.18,330.49) .. (362.14,330.49) .. controls (361.1,330.49) and (360.26,329.65) .. (360.26,328.61) -- cycle ;
\draw   (365.99,307.94) .. controls (367.5,307.94) and (368.72,309.16) .. (368.72,310.66) -- (368.72,329.65) .. controls (368.72,331.15) and (367.5,332.37) .. (365.99,332.37) -- (357.81,332.37) .. controls (356.31,332.37) and (355.09,331.15) .. (355.09,329.65) -- (355.09,310.66) .. controls (355.09,309.16) and (356.31,307.94) .. (357.81,307.94) -- cycle ;

\draw   (412.45,332.13) .. controls (412.45,331.09) and (413.29,330.25) .. (414.33,330.25) .. controls (415.37,330.25) and (416.21,331.09) .. (416.21,332.13) .. controls (416.21,333.16) and (415.37,334.01) .. (414.33,334.01) .. controls (413.29,334.01) and (412.45,333.16) .. (412.45,332.13) -- cycle ;
\draw   (418.18,312.39) .. controls (419.69,312.39) and (420.91,313.61) .. (420.91,315.11) -- (420.91,334.1) .. controls (420.91,335.6) and (419.69,336.82) .. (418.18,336.82) -- (410,336.82) .. controls (408.5,336.82) and (407.28,335.6) .. (407.28,334.1) -- (407.28,315.11) .. controls (407.28,313.61) and (408.5,312.39) .. (410,312.39) -- cycle ;

\path  [shading=_h1f5ktwi3,_6twq0hqa4] (253.31,175.66) .. controls (259.06,191.58) and (310.28,317.65) .. (299.08,324.72) .. controls (287.89,331.8) and (258.29,203.78) .. (252.26,183.08) .. controls (246.23,162.39) and (247.56,159.74) .. (253.31,175.66) -- cycle ; 
 \draw  [color={rgb, 255:red, 0; green, 0; blue, 0 }  ,draw opacity=1 ] (253.31,175.66) .. controls (259.06,191.58) and (310.28,317.65) .. (299.08,324.72) .. controls (287.89,331.8) and (258.29,203.78) .. (252.26,183.08) .. controls (246.23,162.39) and (247.56,159.74) .. (253.31,175.66) -- cycle ; 

\path  [shading=_4i90ziygl,_gl1m40bbq] (340.77,168.38) .. controls (356.7,193.79) and (412.71,292.66) .. (408.01,305.04) .. controls (403.3,317.41) and (346.41,191.84) .. (338.5,174.07) .. controls (330.6,156.31) and (324.84,142.97) .. (340.77,168.38) -- cycle ; 
 \draw  [color={rgb, 255:red, 0; green, 0; blue, 0 }  ,draw opacity=1 ] (340.77,168.38) .. controls (356.7,193.79) and (412.71,292.66) .. (408.01,305.04) .. controls (403.3,317.41) and (346.41,191.84) .. (338.5,174.07) .. controls (330.6,156.31) and (324.84,142.97) .. (340.77,168.38) -- cycle ; 

\draw  [fill={rgb, 255:red, 128; green, 128; blue, 128 }  ,fill opacity=1 ] (376.65,94.76) -- (372.33,92.52) -- (306.04,113.57) -- (315.22,142.49) -- (319.55,144.73) -- (385.84,123.68) -- cycle ; \draw   (306.04,113.57) -- (310.37,115.81) -- (376.65,94.76) ; \draw   (310.37,115.81) -- (319.55,144.73) ;
\draw  [fill={rgb, 255:red, 128; green, 128; blue, 128 }  ,fill opacity=1 ] (271.59,79.98) -- (277.13,81.64) -- (282.68,79.98) -- (356.43,100.87) -- (330.11,109.33) -- cycle ;
\draw  [dash pattern={on 0.84pt off 2.51pt}]  (351.58,117.76) -- (391,81.67) ;
\draw [shift={(349.85,119.35)}, rotate = 317.52] [color={rgb, 255:red, 0; green, 0; blue, 0 }  ][line width=0.75]      (0, 0) circle [x radius= 3.35, y radius= 3.35]   ;
\draw  [dash pattern={on 0.84pt off 2.51pt}]  (259.36,145.39) -- (212,98.67) ;
\draw [shift={(261.03,147.04)}, rotate = 224.61] [color={rgb, 255:red, 0; green, 0; blue, 0 }  ][line width=0.75]      (0, 0) circle [x radius= 3.35, y radius= 3.35]   ;
\path  [shading=_96ealr2rs,_bm6za6y7b] (294.44,145.58) .. controls (305.06,163.72) and (366.01,285.55) .. (359.9,297.3) .. controls (353.79,309.04) and (295.3,161.25) .. (288.69,142.93) .. controls (282.08,124.6) and (283.83,127.45) .. (294.44,145.58) -- cycle ; 
 \draw  [color={rgb, 255:red, 0; green, 0; blue, 0 }  ,draw opacity=1 ] (294.44,145.58) .. controls (305.06,163.72) and (366.01,285.55) .. (359.9,297.3) .. controls (353.79,309.04) and (295.3,161.25) .. (288.69,142.93) .. controls (282.08,124.6) and (283.83,127.45) .. (294.44,145.58) -- cycle ; 

\draw (287.5,22.16) node [anchor=north west][inner sep=0.75pt]   [align=left] {$\displaystyle y$};
\draw (200.55,304.94) node [anchor=north west][inner sep=0.75pt]   [align=left] {$\displaystyle x$};
\draw (463.34,281.19) node [anchor=north west][inner sep=0.75pt]   [align=left] {$\displaystyle z$};
\draw (141.29,74.51) node [anchor=north west][inner sep=0.75pt]  [font=\normalsize] [align=left] {$\displaystyle N_{\text{R}}$-elements RIS};
\draw (343.01,59.38) node [anchor=north west][inner sep=0.75pt]  [font=\normalsize] [align=left] {\begin{minipage}[lt]{82.16pt}\setlength\topsep{0pt}
\begin{flushright}
 $\displaystyle N_{\text{A}}$-antennas BS
\end{flushright}

\end{minipage}};

\end{tikzpicture}

%% file: fig/back_view.tex
\tikzset{every picture/.style={line width=0.75pt}} 

\begin{tikzpicture}[x=0.75pt,y=0.75pt,yscale=-1,xscale=1]

\draw  [draw opacity=0][fill={rgb, 255:red, 155; green, 155; blue, 155 }  ,fill opacity=1 ] (349.33,82.94) -- (358.4,82.92) -- (358.07,97.86) -- (349,97.87) -- cycle ;
\draw  [draw opacity=0][fill={rgb, 255:red, 155; green, 155; blue, 155 }  ,fill opacity=1 ] (357.84,97.94) -- (366.81,97.92) -- (366.83,111.52) -- (357.87,111.54) -- cycle ;
\draw  [draw opacity=0][fill={rgb, 255:red, 155; green, 155; blue, 155 }  ,fill opacity=1 ] (358.95,82.88) -- (368.19,82.86) -- (367,97.92) -- (357.76,97.94) -- cycle ;
\draw  [draw opacity=0][fill={rgb, 255:red, 155; green, 155; blue, 155 }  ,fill opacity=1 ] (263.56,83.57) -- (274.75,83.55) -- (276.25,97.69) -- (265.07,97.71) -- cycle ;
\draw   (368.25,82.9) -- (364,138.55) -- (268.53,139.16) -- (263.56,83.57) -- cycle ;
\draw  [fill={rgb, 255:red, 255; green, 255; blue, 255 }  ,fill opacity=1 ] (227.07,124.12) -- (412.67,124.12) -- (412.67,243.9) -- (227.07,243.9) -- cycle ;
\draw  [color={rgb, 255:red, 155; green, 155; blue, 155 }  ,draw opacity=1 ] (313.3,199.91) .. controls (313.3,197.26) and (315.45,195.11) .. (318.1,195.11) .. controls (320.75,195.11) and (322.9,197.26) .. (322.9,199.91) .. controls (322.9,202.56) and (320.75,204.71) .. (318.1,204.71) .. controls (315.45,204.71) and (313.3,202.56) .. (313.3,199.91) -- cycle ;
\draw [color={rgb, 255:red, 155; green, 155; blue, 155 }  ,draw opacity=1 ] (302,199.91) -- (429.2,199.91)(318.1,60.35) -- (318.1,216.1) (422.2,194.91) -- (429.2,199.91) -- (422.2,204.91) (313.1,67.35) -- (318.1,60.35) -- (323.1,67.35)  ;
\draw    (275.2,83.72) -- (278.2,124.6) ;
\draw    (286.8,83.32) -- (289.8,124.6) ;
\draw    (298,83.32) -- (300.6,123.8) ;
\draw    (309.6,83.72) -- (311.4,124.2) ;
\draw    (326.4,83.72) -- (325.4,123.8) ;
\draw    (337.6,83.32) -- (337,123.8) ;
\draw    (349.2,82.92) -- (347.8,123.8) ;
\draw    (358.4,82.92) -- (357,123.8) ;
\draw    (264.8,97.92) -- (367,97.92) ;
\draw    (266.2,111.8) -- (366.6,112.2) ;

\draw [color={rgb, 255:red, 128; green, 128; blue, 128 }  ,draw opacity=1 ][line width=1.5]    (262.85,79.8) -- (271.65,79.8) ;
\draw [shift={(275.65,79.8)}, rotate = 180] [fill={rgb, 255:red, 128; green, 128; blue, 128 }  ,fill opacity=1 ][line width=0.08]  [draw opacity=0] (4.64,-2.23) -- (0,0) -- (4.64,2.23) -- cycle    ;
\draw [color={rgb, 255:red, 128; green, 128; blue, 128 }  ,draw opacity=1 ][line width=1.5]    (260,83.2) -- (260.94,95.21) ;
\draw [shift={(261.25,99.2)}, rotate = 265.53] [fill={rgb, 255:red, 128; green, 128; blue, 128 }  ,fill opacity=1 ][line width=0.08]  [draw opacity=0] (4.64,-2.23) -- (0,0) -- (4.64,2.23) -- cycle    ;
\draw [color={rgb, 255:red, 128; green, 128; blue, 128 }  ,draw opacity=1 ][line width=1.5]    (223.5,123.7) -- (223.5,141.2) ;
\draw [shift={(223.5,145.2)}, rotate = 270] [fill={rgb, 255:red, 128; green, 128; blue, 128 }  ,fill opacity=1 ][line width=0.08]  [draw opacity=0] (4.64,-2.23) -- (0,0) -- (4.64,2.23) -- cycle    ;
\draw [color={rgb, 255:red, 128; green, 128; blue, 128 }  ,draw opacity=1 ][line width=1.5]    (226.9,120.37) -- (245,120.37) ;
\draw [shift={(249,120.37)}, rotate = 180] [fill={rgb, 255:red, 128; green, 128; blue, 128 }  ,fill opacity=1 ][line width=0.08]  [draw opacity=0] (4.64,-2.23) -- (0,0) -- (4.64,2.23) -- cycle    ;
\draw  [draw opacity=0][fill={rgb, 255:red, 128; green, 128; blue, 128 }  ,fill opacity=1 ] (229.4,126.12) -- (250,126.12) -- (250,145.7) -- (229.4,145.7) -- cycle ;
\draw  [draw opacity=0][fill={rgb, 255:red, 128; green, 128; blue, 128 }  ,fill opacity=1 ] (229.4,146.7) -- (250,146.7) -- (250,166.28) -- (229.4,166.28) -- cycle ;
\draw  [draw opacity=0][fill={rgb, 255:red, 128; green, 128; blue, 128 }  ,fill opacity=1 ] (251,126.12) -- (271.6,126.12) -- (271.6,145.7) -- (251,145.7) -- cycle ;
\draw  [draw opacity=0][fill={rgb, 255:red, 128; green, 128; blue, 128 }  ,fill opacity=1 ] (250.73,146.79) -- (271.6,146.79) -- (271.33,166.37) -- (250.73,166.37) -- cycle ;
\draw [color={rgb, 255:red, 128; green, 128; blue, 128 }  ,draw opacity=1 ]   (275.3,134.91) -- (275.2,155.58) ;
\draw [shift={(275.2,155.58)}, rotate = 270.28] [color={rgb, 255:red, 128; green, 128; blue, 128 }  ,draw opacity=1 ][line width=0.75]    (0,2.24) -- (0,-2.24)   ;
\draw [shift={(275.3,134.91)}, rotate = 270.28] [color={rgb, 255:red, 128; green, 128; blue, 128 }  ,draw opacity=1 ][line width=0.75]    (0,2.24) -- (0,-2.24)   ;
\draw [color={rgb, 255:red, 128; green, 128; blue, 128 }  ,draw opacity=1 ]   (238.7,170.49) -- (260.2,170.58) ;
\draw [shift={(260.2,170.58)}, rotate = 180.24] [color={rgb, 255:red, 128; green, 128; blue, 128 }  ,draw opacity=1 ][line width=0.75]    (0,2.24) -- (0,-2.24)   ;
\draw [shift={(238.7,170.49)}, rotate = 180.24] [color={rgb, 255:red, 128; green, 128; blue, 128 }  ,draw opacity=1 ][line width=0.75]    (0,2.24) -- (0,-2.24)   ;
\draw [color={rgb, 255:red, 128; green, 128; blue, 128 }  ,draw opacity=1 ]   (371.63,89.91) -- (370,107.53) ;
\draw [shift={(370,107.53)}, rotate = 275.3] [color={rgb, 255:red, 128; green, 128; blue, 128 }  ,draw opacity=1 ][line width=0.75]    (0,2.24) -- (0,-2.24)   ;
\draw [shift={(371.63,89.91)}, rotate = 275.3] [color={rgb, 255:red, 128; green, 128; blue, 128 }  ,draw opacity=1 ][line width=0.75]    (0,2.24) -- (0,-2.24)   ;
\draw [color={rgb, 255:red, 128; green, 128; blue, 128 }  ,draw opacity=1 ]   (353.37,78.16) -- (364.67,78.16) ;
\draw [shift={(364.67,78.16)}, rotate = 180] [color={rgb, 255:red, 128; green, 128; blue, 128 }  ,draw opacity=1 ][line width=0.75]    (0,2.24) -- (0,-2.24)   ;
\draw [shift={(353.37,78.16)}, rotate = 180] [color={rgb, 255:red, 128; green, 128; blue, 128 }  ,draw opacity=1 ][line width=0.75]    (0,2.24) -- (0,-2.24)   ;

\draw (302,203) node [anchor=north west][inner sep=0.75pt]  [color={rgb, 255:red, 155; green, 155; blue, 155 }  ,opacity=1 ] [align=left] {$\displaystyle z$};
\draw (427,202) node [anchor=north west][inner sep=0.75pt]  [color={rgb, 255:red, 155; green, 155; blue, 155 }  ,opacity=1 ] [align=left] {$\displaystyle x$};
\draw (299.7,56.1) node [anchor=north west][inner sep=0.75pt]  [color={rgb, 255:red, 155; green, 155; blue, 155 }  ,opacity=1 ] [align=left] {$\displaystyle y$};
\draw (231,223) node [anchor=north west][inner sep=0.75pt]  [font=\small] [align=left] {$\displaystyle N_{\text{R}}$-elements RIS};
\draw (255.73,38.12) node [anchor=north west][inner sep=0.75pt]  [font=\small] [align=left] {\begin{minipage}[lt]{82.89pt}\setlength\topsep{0pt}
\begin{flushright}
 $\displaystyle N_{\text{A}}$-antennas array
\end{flushright}

\end{minipage}};
\draw (265.7,58.6) node [anchor=north west][inner sep=0.75pt]  [font=\small,color={rgb, 255:red, 0; green, 0; blue, 0 }  ,opacity=1 ] [align=left] {$\displaystyle \Delta _{\text{A}}^{\text{h}}$};
\draw (237.7,79.1) node [anchor=north west][inner sep=0.75pt]  [font=\small,color={rgb, 255:red, 0; green, 0; blue, 0 }  ,opacity=1 ] [align=left] {$\displaystyle \Delta _{\text{A}}^{\text{v}}$};
\draw (215,98) node [anchor=north west][inner sep=0.75pt]  [font=\small,color={rgb, 255:red, 0; green, 0; blue, 0 }  ,opacity=1 ] [align=left] {$\displaystyle \Delta _{\text{R}}^{\text{h}}$};
\draw (198.7,124.6) node [anchor=north west][inner sep=0.75pt]  [font=\small,color={rgb, 255:red, 0; green, 0; blue, 0 }  ,opacity=1 ] [align=left] {$\displaystyle \Delta _{\text{R}}^{\text{v}}$};
\draw (240.7,173.49) node [anchor=north west][inner sep=0.75pt]  [font=\small,color={rgb, 255:red, 0; green, 0; blue, 0 }  ,opacity=1 ] [align=left] {$\displaystyle d_{\text{R}}^{\text{h}}$};
\draw (278.7,133.93) node [anchor=north west][inner sep=0.75pt]  [font=\small,color={rgb, 255:red, 0; green, 0; blue, 0 }  ,opacity=1 ] [align=left] {$\displaystyle d_{\text{R}}^{\text{v}}$};
\draw (376.03,88.6) node [anchor=north west][inner sep=0.75pt]  [font=\small,color={rgb, 255:red, 0; green, 0; blue, 0 }  ,opacity=1 ] [align=left] {$\displaystyle d_{\text{A}}^{\text{v}}$};
\draw (358.37,57.16) node [anchor=north west][inner sep=0.75pt]  [font=\small,color={rgb, 255:red, 0; green, 0; blue, 0 }  ,opacity=1 ] [align=left] {$\displaystyle d_{\text{A}}^{\text{h}}$};

\end{tikzpicture}

%% file: fig/side_view.tex
\tikzset{every picture/.style={line width=0.75pt}} 

\begin{tikzpicture}[x=0.75pt,y=0.75pt,yscale=-1,xscale=1]

\draw  (192,236.47) -- (414.8,236.47)(241.8,49) -- (241.8,249.47) (407.8,231.47) -- (414.8,236.47) -- (407.8,241.47) (236.8,56) -- (241.8,49) -- (246.8,56)  ;
\draw  [fill={rgb, 255:red, 128; green, 128; blue, 128 }  ,fill opacity=1 ] (239.8,100) -- (244.8,100) -- (244.8,169.47) -- (239.8,169.47) -- cycle ;
\draw [color={rgb, 255:red, 155; green, 155; blue, 155 }  ,draw opacity=1 ] [dash pattern={on 4.5pt off 4.5pt}]  (193.33,99.28) -- (193.33,235) ;
\draw [shift={(193.33,99.28)}, rotate = 270] [color={rgb, 255:red, 155; green, 155; blue, 155 }  ,draw opacity=1 ][line width=0.75]    (0,5.59) -- (0,-5.59)   ;
\draw [color={rgb, 255:red, 155; green, 155; blue, 155 }  ,draw opacity=1 ] [dash pattern={on 0.84pt off 2.51pt}]  (244.8,100) -- (301.27,100) ;
\draw [color={rgb, 255:red, 155; green, 155; blue, 155 }  ,draw opacity=1 ] [dash pattern={on 0.84pt off 2.51pt}]  (308.27,67.88) -- (308.27,106.67) ;
\draw  [fill={rgb, 255:red, 128; green, 128; blue, 128 }  ,fill opacity=1 ] (292.3,88.3) -- (296.1,85.64) -- (313.69,110.67) -- (309.89,113.34) -- cycle ;
\draw [color={rgb, 255:red, 155; green, 155; blue, 155 }  ,draw opacity=1 ] [dash pattern={on 4.5pt off 4.5pt}]  (213.27,133.67) -- (213.27,236.21) ;
\draw [shift={(213.27,133.67)}, rotate = 270] [color={rgb, 255:red, 155; green, 155; blue, 155 }  ,draw opacity=1 ][line width=0.75]    (0,5.59) -- (0,-5.59)   ;
\draw   (356.47,192.57) .. controls (356.47,191.46) and (357.36,190.57) .. (358.47,190.57) .. controls (359.57,190.57) and (360.47,191.46) .. (360.47,192.57) .. controls (360.47,193.67) and (359.57,194.57) .. (358.47,194.57) .. controls (357.36,194.57) and (356.47,193.67) .. (356.47,192.57) -- cycle ;
\draw   (362.57,171.57) .. controls (364.17,171.57) and (365.47,172.87) .. (365.47,174.47) -- (365.47,194.67) .. controls (365.47,196.27) and (364.17,197.57) .. (362.57,197.57) -- (353.87,197.57) .. controls (352.27,197.57) and (350.97,196.27) .. (350.97,194.67) -- (350.97,174.47) .. controls (350.97,172.87) and (352.27,171.57) .. (353.87,171.57) -- cycle ;
\draw [color={rgb, 255:red, 155; green, 155; blue, 155 }  ,draw opacity=1 ] [dash pattern={on 4.5pt off 4.5pt}]  (383.27,184.48) -- (383.27,236.61) ;
\draw [shift={(383.27,184.48)}, rotate = 270] [color={rgb, 255:red, 155; green, 155; blue, 155 }  ,draw opacity=1 ][line width=0.75]    (0,5.59) -- (0,-5.59)   ;
\draw  [dash pattern={on 4.5pt off 4.5pt}]  (252.68,137.28) .. controls (291.88,162.08) and (314.68,159.28) .. (341.88,181.67) ;
\draw [line width=1.5]  [dash pattern={on 5.63pt off 4.5pt}]  (252,126.31) .. controls (269,115.48) and (281,109.48) .. (296.48,103.68) ;
\draw  [draw opacity=0] (300.52,91.74) .. controls (300.81,89.76) and (301.76,88.14) .. (303.4,87.18) .. controls (304.82,86.34) and (306.59,86.09) .. (308.5,86.36) -- (312.79,98.98) -- cycle ; \draw   (300.52,91.74) .. controls (300.81,89.76) and (301.76,88.14) .. (303.4,87.18) .. controls (304.82,86.34) and (306.59,86.09) .. (308.5,86.36) ;  
\draw [color={rgb, 255:red, 155; green, 155; blue, 155 }  ,draw opacity=1 ] [dash pattern={on 4.5pt off 4.5pt}]  (308.27,63.88) -- (241.2,63.88) ;
\draw [shift={(308.27,63.88)}, rotate = 360] [color={rgb, 255:red, 155; green, 155; blue, 155 }  ,draw opacity=1 ][line width=0.75]    (0,5.59) -- (0,-5.59)   ;
\draw   (237,236.47) .. controls (237,233.82) and (239.15,231.67) .. (241.8,231.67) .. controls (244.45,231.67) and (246.6,233.82) .. (246.6,236.47) .. controls (246.6,239.13) and (244.45,241.27) .. (241.8,241.27) .. controls (239.15,241.27) and (237,239.13) .. (237,236.47) -- cycle ;

\draw (328,202) node [anchor=north west][inner sep=0.75pt]  [font=\normalsize] [align=left] {$\displaystyle k$-th UE};
\draw (373,161) node [anchor=north west][inner sep=0.75pt]  [color={rgb, 255:red, 128; green, 128; blue, 128 }  ,opacity=1 ] [align=left] {$\displaystyle h_{\text{UE}}$};
\draw (182,74) node [anchor=north west][inner sep=0.75pt]  [color={rgb, 255:red, 128; green, 128; blue, 128 }  ,opacity=1 ] [align=left] {$\displaystyle h_{\text{BS}}$};
\draw (201,111) node [anchor=north west][inner sep=0.75pt]  [color={rgb, 255:red, 128; green, 128; blue, 128 }  ,opacity=1 ] [align=left] {$\displaystyle h_{\text{RIS}}$};
\draw (295,70) node [anchor=north west][inner sep=0.75pt]   [align=left] {$\displaystyle \alpha $};
\draw (277,114) node [anchor=north west][inner sep=0.75pt]   [align=left] {$\displaystyle \mathbf{H}$};
\draw (313,141) node [anchor=north west][inner sep=0.75pt]   [align=left] {$\displaystyle \mathbf{h}_{k}$};
\draw (416,236) node [anchor=north west][inner sep=0.75pt]   [align=left] {$\displaystyle z$};
\draw (222,37) node [anchor=north west][inner sep=0.75pt]   [align=left] {$\displaystyle y$};
\draw (271,44) node [anchor=north west][inner sep=0.75pt]  [color={rgb, 255:red, 128; green, 128; blue, 128 }  ,opacity=1 ] [align=left] {$\displaystyle D$};
\draw (248,155) node [anchor=north west][inner sep=0.75pt]   [align=left] {$\displaystyle \mathbf{P}$};
\draw    (333.8,77) -- (403.8,77) -- (403.8,135.6) -- (333.8,135.6) -- cycle  ;
\draw (400.8,131.6) node [anchor=south east] [inner sep=0.75pt]  [font=\normalsize] [align=left] {\begin{minipage}[lt]{44.57pt}\setlength\topsep{0pt}
\begin{center}
{\normalsize cascaded }\\{\normalsize channel}\\$\displaystyle \mathbf{HPh}_{k}$
\end{center}

\end{minipage}};
\draw (226,242) node [anchor=north west][inner sep=0.75pt]   [align=left] {$\displaystyle x$};

\end{tikzpicture}